\documentclass[aps,prc,superscriptaddress,reprint,nofootinbib]{revtex4-1}
\pdfoutput=1

\usepackage{epsfig}
\usepackage{dcolumn}
\usepackage{bm}
\usepackage{amssymb}
\usepackage{amsmath}
\usepackage{hyperref}

\usepackage{graphicx}
\usepackage[dvipsnames]{xcolor}
\usepackage{color}
\definecolor{darkblue}{rgb}{0.0, 0.0, 0.55}
\definecolor{cite}{rgb}{0.0, 0.34, 0.25}
\definecolor{midgreen}{rgb}{0.52, 0.73, 0.4}

\usepackage[utf8]{inputenc}

\usepackage{rotating}
\usepackage{bigints}

\newcommand{\MeV}{\, {\rm MeV}}

\begin{document}
\title{Strangeness neutral equation of state for QCD with a critical point}

\author{J. M. Stafford}
\email[Corresponding author: ]{jmstafford@uh.edu}
\affiliation{Department of Physics, University of Houston, Houston, TX 77204, USA}

\author{D. Mroczek}
\affiliation{Department of Physics, 
University of Illinois at Urbana-Champaign, Urbana, IL 61801, USA}

\author{A. R. Nava Acuna}
\affiliation{Department of Physics, University of Houston, Houston, TX 77204, USA}

\author{J. Noronha-Hostler}
\affiliation{Department of Physics, 
University of Illinois at Urbana-Champaign, Urbana, IL 61801, USA}

\author{P. Parotto}
\affiliation{University of Wuppertal, Department of Physics, Wuppertal D-42119, Germany}

\author{D. R. P. Price}
\affiliation{Department of Physics, University of Houston, Houston, TX 77204, USA}

\author{C. Ratti}
\affiliation{Department of Physics, University of Houston, Houston, TX 77204, USA}

\date{\today}

\begin{abstract}
We present a strangeness-neutral equation of state for QCD that exhibits critical behavior and matches lattice QCD results for the Taylor-expanded thermodynamic variables up to fourth-order in $\mu_B/T$. It is compatible with the SMASH hadronic transport approach and has a range of temperatures and baryonic chemical potentials relevant for phase II of the Beam Energy Scan at RHIC.  We provide an updated version of the software BES-EoS, which produces an equation of state for QCD that includes a critical point in the 3D Ising model universality class. This new version also includes isentropic trejectories and the critical contribution to the correlation length. 
Since heavy-ion collisions have  zero global net-strangeness density and a fixed ratio of electric charge to baryon number, the BES-EoS is more suitable to describe this system.
Comparison with the previous version of the EoS is thoroughly discussed.
\end{abstract}

\maketitle

\section{Introduction}
\label{sec:intro}
One of the main open questions in hot and dense strongly interacting matter is whether a critical point on the QCD phase diagram separates the established crossover at small baryonic density \cite{Aoki:2006we,Borsanyi:2010bp,Endrodi:2011gv,Bellwied:2015rza,Guenther:2020vqg} from a first order phase transition, hypothesized to exist as the asymmetry between matter and anti-matter gets increasingly large \cite{Stephanov:2011pb,Ratti:2006gh,Critelli:2017oub,Ratti:2018ksb,Bzdak:2019pkr}. While the fermionic sign problem currently prevents a definitive answer from first principles lattice QCD simulations, the critical point search is at the core of the second Beam Energy Scan (BES-II) at the Relativistic Heavy Ion Collider (RHIC) in Brookhaven, which will take data until 2021.

The equation of state (EoS) of QCD plays a crucial role in the search for the critical point, and it is needed as input in the hydrodynamic simulations that describe the system created in the collisions \cite{Dexheimer:2020zzs,Monnai:2021kgu}. The knowledge of the QCD equation of state and phase diagram at finite density from first principles is currently limited,  due to the aforementioned sign problem. The EoS at vanishing chemical potential $\mu_B=0$
is available over a broad range of temperatures
\cite{Borsanyi:2010cj,Borsanyi:2013bia,Bazavov:2014pvz}.
The first continuum-extrapolated extension to finite $\mu_B$ in Ref.~\cite{Borsanyi:2012cr}, based on the Taylor expansion method, was followed by
other works in which more terms were added to the Taylor series, with the intent of extending the $\mu_B$-range to higher values
\cite{Bazavov:2017dus,Gunther:2016vcp,Bellwied:2015lba,Gunther:2017sxn,Borsanyi:2018grb,Bazavov:2020bjn}.  Recently, a new expansion scheme was proposed in Ref. \cite{Borsanyi:2021sxv}, which shows improved convergence properties compared to the Taylor series. Results on the EoS at finite $\mu_B,~\mu_S$ and $\mu_Q$
are also available \cite{Noronha-Hostler:2019ayj,Monnai:2019hkn}.

To support the RHIC experimental effort, theoretical predictions on the location of the critical point \cite{Critelli:2017oub,Bazavov:2017dus,DElia:2016jqh,Pasztor:2018yae,Giordano:2020roi,Fodor:2004nz,Datta:2016ukp,Fischer:2012vc,Eichmann:2015kfa,Borsanyi:2020fev} and its effect on observables \cite{An:2020vri,Mroczek:2020rpm,Bluhm:2020mpc,Nahrgang:2020yxm,Nahrgang:2018afz,Sorensen:2020ygf} are steadily becoming available, together with approaches to extend the reach of the QCD equation of state to high density \cite{Grefa:2021qvt,Motornenko:2019arp,Vovchenko:2018zgt}. In this context, some of us have recently developed a family of equations of state, which reproduce the lattice QCD one in the density regime where it is available, and contain a critical point in the 3D Ising model universality class (the one expected for QCD) \cite{Parotto:2018pwx}.
In this approach, the location of the critical point and the strength of the critical region can be chosen at will, and can then be constrained through comparisons with the experimental data from BES-II. Other works have already used this new EoS to study out-of-equilibrium approaches to the QCD critical point \cite{Dore:2020jye,Hama:2020nfp}, the sign of the kurtosis \cite{Mroczek:2020rpm}, and to calculate transport coefficients \cite{McLaughlin:2021dph}. 

In this manuscript, we improve the results presented in Ref. \cite{Parotto:2018pwx} by introducing the conditions of strangeness neutrality and electric charge conservation into our EoS, in order to match the experimental situation in a heavy-ion collision. The hadron resonance gas (HRG) model is used to smoothly continue the lattice QCD results to small temperatures. The resonance list on which the hadronic equation of state is based is consistent with the list of particles currently used in the SMASH \cite{Petersen:2018jag} code, so that our EoS can be used in hydrodynamic simulations of heavy-ion collisions (HICs) that are coupled to the SMASH hadronic transport code. We thoroughly compare the strangeness-neutral EoS to the previous version. 
The isentropic trajectories showing the path of the system through the phase diagram are markedly affected by the constraints on the conserved charges. Finally, our updated EoS code outputs the critical contribution to the correlation length for the first time.

\section{The Scaling Equation of State}\label{sec:scal_EoS}
This work is based upon a previously established equation of state (EoS) that incorporates critical behavior, developed within the framework of the BEST collaboration in Ref. \cite{Parotto:2018pwx}. In our updated version \cite{BESTrepos}, we impose the conditions of strangeness neutrality and compatibility with the hadronic transport simulator SMASH \cite{Petersen:2018jag}, by using the SMASH hadronic list as input in the HRG model. 
Because SMASH has become a standard transport code used within the field, we ensure consistency across all stages of phenomenological modeling of heavy-ion collisions. We begin by describing the general procedure for developing an EoS with a critical point in the 3D Ising model universality class. We then provide details of the implementation of the new features into the EoS.

In order to study the effect of a critical point that could potentially be observed during the BES-II at RHIC on QCD thermodynamics, we utilize the 3D Ising model to map such critical behavior onto the phase diagram of QCD. The 3D Ising model was chosen for this approach because it exhibits the same scaling features in the vicinity of a critical point as QCD, in other words they belong to the same universality class \cite{Pisarski:1983ms,Rajagopal:1992qz}. We implement the non-universal mapping of the 3D Ising model onto the QCD phase diagram in such a way that the Taylor expansion coefficients of our final pressure match the ones calculated on the lattice order by order. This prescription can be summarized as follows:
\begin{enumerate}
    \item Define a parametrization of the 3D Ising model near the critical point, consistent with what has been previously shown in the literature  \cite{Guida:1996ep, Nonaka:2004pg, Stephanov:2011pb, Parotto:2018pwx}, 
    \begin{equation} \label{IsingEoS}
        \begin{split}
            M &= M_0 R^{\beta} \theta \\
            h &= h_0 R^{\beta \delta} \tilde{h}(\theta) \\
            r &= R(1- \theta^2)
        \end{split}
    \end{equation}
	where the magnetization $M$, the  magnetic field $h$, and the reduced temperature $r$, are given in terms of the external parameters $R$ and $\theta$. The normalization constants for the magnetization and magnetic field are $M_0=0.605$ and $h_0=0.364$, respectively, $\beta=0.326$ and $\delta=4.8$ are critical exponents in the 3D Ising Model, and $\tilde{h}(\theta$)=$\theta$ (1-0.76201$\theta^2$+0.00804$\theta^4$).
	
	The singular part of the pressure is described by the parametrized Gibbs' free energy:
	\begin{equation}
	\label{GFreeEner}
	    \begin{split}
	     P_{\text{Ising}} &= - G(R,\theta) \\
	     &= h_0 M_0 R^{2 - \alpha}(\theta \tilde{h}(\theta) - g(\theta)),
	    \end{split}
	\end{equation}
	where
	\begin{align*}
	\centering
	     g(\theta) &= c_0 +  c_1(1-\theta^2) + c_2(1-\theta^2)^2 + c_3(1-\theta^2)^3, \\
        c_0 &= \frac{\beta}{2-\alpha}(1+a+b), \\
	 c_1 &= -\frac{1}{2} \frac{1}{\alpha -1}((1-2\beta)(1+a+b)-2\beta(a+2b)), \\
	 c_2 &= - \frac{1}{2\alpha}(2\beta b - (1-2\beta)(a+2b)), \\
	 c_3 &= - \frac{1}{2(\alpha+1)}b(1-2\beta).
    \end{align*}
	 
	Because QCD is symmetric about $\mu_B=0$, we require that  $P_{\rm{Ising}}$ is also matter-anti-matter symmetric. Thus, we perform the calculations in a range of $\mu_B$ spanning positive and negative values. Furthermore, the equations defined here are subject to the following constraints on the parameters: R $\geq$ 0, $\lvert \theta \rvert$ $\leq$ $\theta_0$ $\sim$ 1.154.
    \item Choose the location of the critical point and map the critical behavior onto the QCD phase diagram via a linear map from \{$T$, $\mu_B$\} to \{$r,h$\}: 
    \begin{equation} \label{mapT}
        \frac{T-T_c}{T_c}=\omega(\rho r \sin{\alpha_1} + h \sin{\alpha_2})
    \end{equation}
    \begin{equation} \label{mapmuB}
        \frac{\mu_B-\mu_{B,c}}{T_c} = \omega(-\rho r \cos{\alpha_1} - h \cos{\alpha_2})
    \end{equation}
    
    where $(T_c , \mu_{B,c})$ are the coordinates of the critical point, and ($\alpha_1$,$\alpha_2$) are the angles between the axes of the QCD phase diagram and the Ising model ones. Finally, $\omega$ and $\rho$ are scaling parameters for the Ising-to-QCD map: $\omega$ determines the overall scale of both $r$ and $h$, while $\rho$ determines the relative scale between them.

    \item As previously established in Ref.~\cite{Parotto:2018pwx}, we reduce the number of free parameters from six to four, by assuming the critical point sits on the chiral phase transition line, and by imposing that the $r$ axis of the Ising model is tangent to the transition line of QCD at the critical point: 
    \begin{equation} \label{chiraltrans}
        T=T_0 + \kappa \, T_0 \, \left(\frac{\mu_B}{T_0}\right)^2 + \mathcal{O}(\mu_B^4).
    \end{equation}
    In this study, we maintain consistency with the original EoS development of Ref.~\cite{Parotto:2018pwx} by utilizing the same parameters\footnote{As in Ref.~\cite{Parotto:2018pwx}, we assume the transition line to be a parabola, and utilize the curvature parameter $\kappa=-0.0149$ from Ref.~\cite{Bellwied:2015rza}. Recent results from lattice QCD \cite{Bazavov:2018mes,Borsanyi:2020fev} are consistent with this value, and predict the next to leading order parameter $\kappa_4$ to be consistent with 0 within errors.}. The critical point lies at \{$T_c \simeq$ 143.2 MeV, $\mu_{B,c}$=350 MeV\}, while the angular parameters are orthogonal $\alpha_1$=3.85\textdegree and $\alpha_2$=93.85\textdegree, and the scaling parameters are $\omega$=1 and $\rho$=2.  However, we remind the reader that such a choice of parameters only has an illustrative purpose, and that we do not make any statement about the position of the critical point or the size of the critical region. As this framework does not serve to yield a prediction for the critical point, but rather to provide an estimate of the effect of critical features on heavy-ion-collision systems, the users can pick their preferred choice of the parameters and test its effect on observables. In particular, we note that by varying the parameters $\omega$ and $\rho$ it is possible to  increase or decrease the effects of the critical point \cite{Parotto:2018pwx,Bzdak:2019pkr}.  Hopefully, experimental data from the BES-II will allow us to constrain the parameters and narrow down the location of the critical point. 
    
    \item Calculate the Ising model susceptibilities and match the Taylor expansion coefficients order by order to Lattice QCD results at $\mu_B$=0. The Taylor expansion of the pressure in $\mu_B$/T as calculated on the lattice can be written as:
    \begin{equation} \label{pressTaylor}
    \begin{split}
        \frac{P(T,\mu_B)}{T^4} = \sum_n c_{2n}(T) \left(\frac{\mu_B}{T} \right)^{2n}.
    \end{split}
    \end{equation}
    Thus, the background pressure, or the non-Ising pressure is, by construction, the difference between the lattice and Ising contributions:
    \begin{equation} \label{coeffmatch}
        T^4 c_{2n}^{\rm{LAT}}(T) = T^4 c_{2n}^{\rm{Non-Ising}}(T) + T_c^4 c_{2n}^{\rm{Ising}}(T).
    \end{equation}
    \item Reconstruct the full Taylor-expanded pressure, including its critical and non-critical components
    \begin{equation} \label{fullpress}
    \begin{split}
        P(T,\mu_B)=T^4 \sum_n c_{2n}^{\rm{Non-Ising}}(T) \left(\frac{\mu_B}{T} \right)^{2n}
        \\
        + P_{\rm{crit}}^{\rm{QCD}}(T,\mu_B),
    \end{split}
    \end{equation}
    where $P_{\rm{crit}}^{\rm{QCD}}$ is the critical contribution to the pressure that has been mapped onto the QCD phase diagram as described in steps 1 and 2.
    
    \item Merge the full reconstructed pressure from the previous step with the HRG pressure at low temperature in order to smooth any non-physical artifacts of the Taylor expansion. For this smooth merging we utilize the hyperbolic tangent:
    \begin{equation} \label{eq:Pmerging}
    \begin{split}
        \frac{P_{\text{Final}}(T,\mu_B)}{T^4} = \frac{P(T,\mu_B)}{T^4} \frac{1}{2}
        \Big[1 + \tanh{\Big(\frac{T-T'(\mu_B)}{\Delta T}}\Big)\Big] \\
        + \frac{P_{HRG}(T,\mu_B)}{T^4} \frac{1}{2}
        \Big[1 - \tanh{\Big(\frac{T-T'(\mu_B)}{\Delta T}}\Big)\Big],
    \end{split}
\end{equation}
    where $T'(\mu_B)$ acts as the switching temperature and $\Delta T$ is the overlap region where both terms contribute. We perform the same merging as in the original development of the EoS, and therefore, perform the merging parallel the QCD transition line and with an overlap region of $\Delta T$=17 MeV.
    \item Calculate thermodynamic quantities as derivatives of the pressure.
\end{enumerate}
For a thorough description of this procedure including investigation of the parameter space and further discussion, we refer the reader to the original development of this framework in Ref. \cite{Parotto:2018pwx}
\section{Strangeness neutrality and compatibility with SMASH}\label{sec:SN_SMASH}

\begin{figure*}[t]
    \centering
    \begin{tabular}{c c}
    \includegraphics[width=0.5\textwidth]{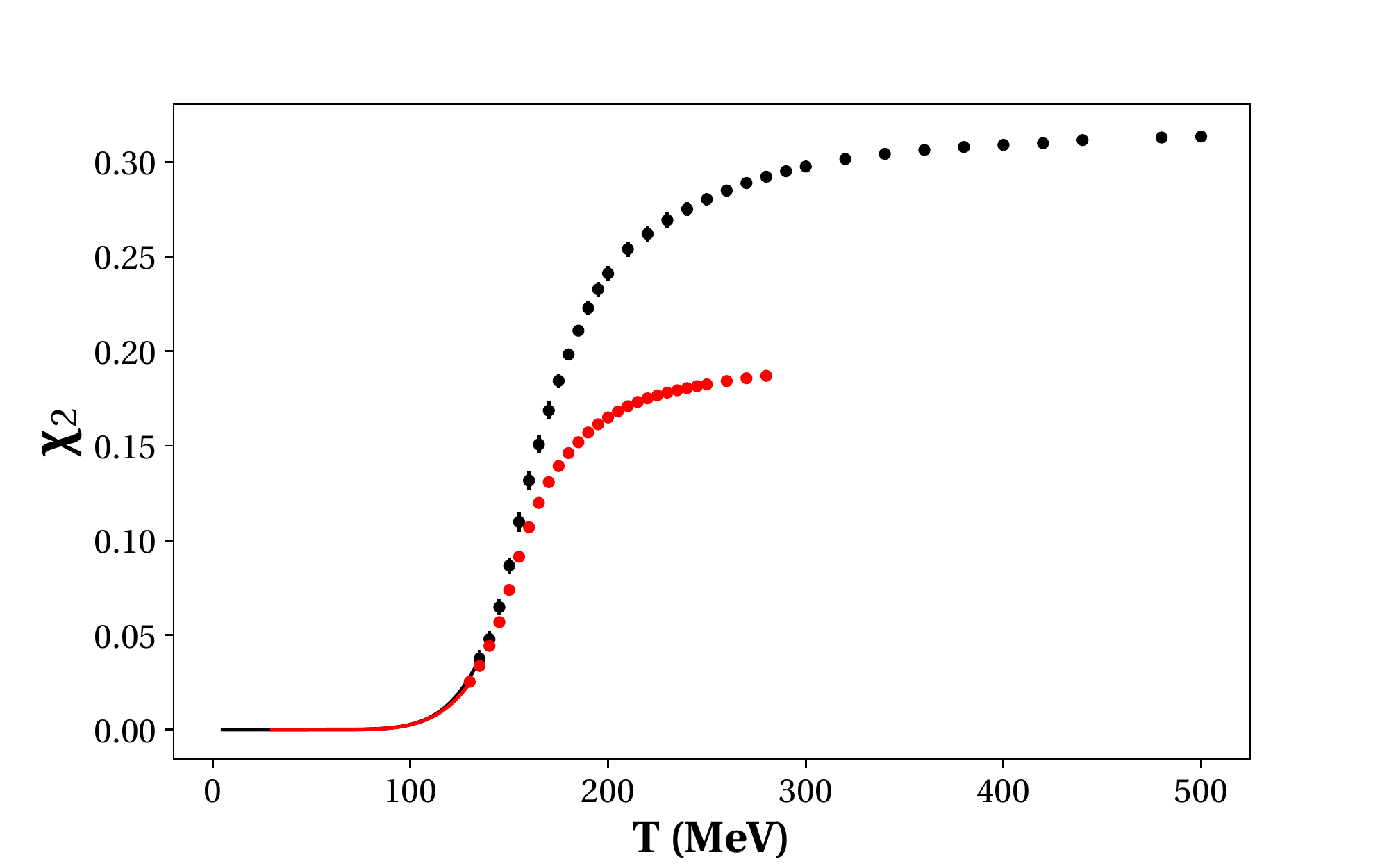} &
    \includegraphics[width=0.5\textwidth]{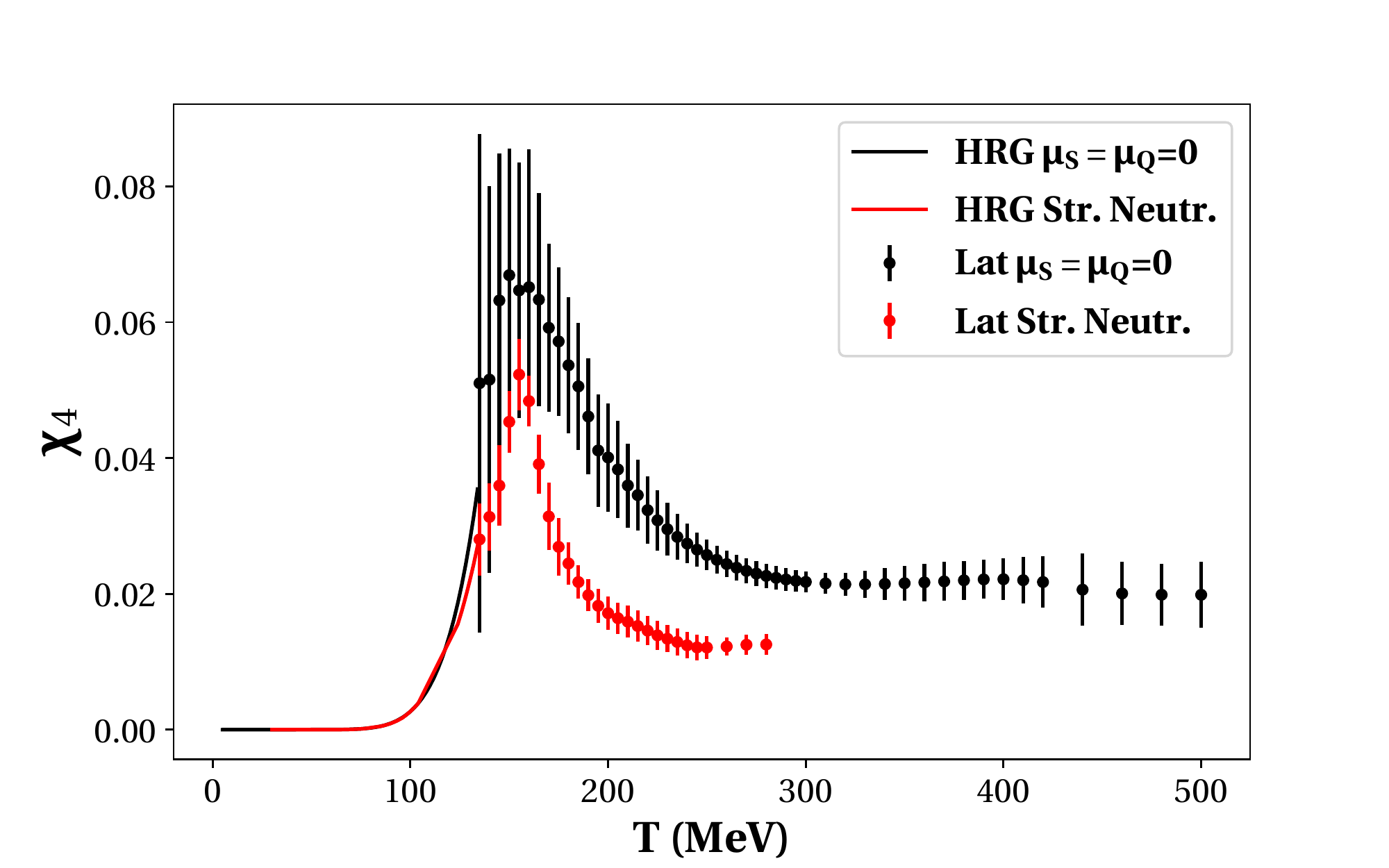}
    \end{tabular}
    \caption{Results for the Taylor expansion coefficients of the pressure from lattice QCD, merged with HRG model calculations at low temperature.  The red points and lines correspond to strangeness neutrality, whereas the black points and lines show the $\mu_S=\mu_Q=0$ case \cite{Bellwied:2015rza, Bellwied:2015lba}. }
    \label{fig:comp_latt_coeff}
\end{figure*}

This updated version of the code that calculates an EoS for QCD via the prescription described in Sec. \ref{sec:scal_EoS} now allows the user to choose between a new, strangeness neutral EoS or the previous one, which is given at $\mu_S$=$\mu_Q$=0. One of the advantages of the strangeness neutral EoS is that the relationships between the conserved charges of QCD mirror those present in a heavy-ion collision. These conditions are incorporated by enforcing a set of requirements on the densities as shown in Eqs. \eqref{strneu}.
\begin{equation} \label{strneu}
\begin{split}
\langle n_S \rangle &= 0 \\
\langle n_Q \rangle &= 0.4 \langle n_B \rangle.
\end{split}
\end{equation}
Given these relations, it is clear that the calculations performed in Lattice QCD and in the HRG model will be different in the case of strangeness neutrality versus $\mu_S$=$\mu_Q$=0. Results on the Taylor coefficients calculated in Lattice QCD are available in both cases \cite{Bellwied:2015rza, Bellwied:2015lba}. Figure \ref{fig:comp_latt_coeff} shows the comparison of the quantities $\chi_2$ and $\chi_4$ from the lattice for both scenarios. While in either case the Taylor coefficients exhibit the expected qualitative features, they are quite different quantitatively, even approaching separate Stefan-Boltzmann limits. Generally, the strangeness neutral trajectory leads to smaller susceptibilities at high temperatures.  Additionally, the peak in $\chi_4$ appears to be narrower for the strangeness neutral line.

The new EoS that obeys the charge conservation conditions present in HICs has the lattice Taylor coefficients calculated under those same conditions as its backbone. Additionally, in order to provide an equation of state in the temperature range required for hydrodynamic simulations, we must also employ the pressure from the HRG model, such that the lattice coefficients can be extended to low enough temperatures and to obtain a smooth parametrization of each term in the Taylor expansion. Therefore, we also implement strangeness neutrality conditions on our HRG model calculations by finding the dependence of the strangeness and electric charge chemical potentials on $T$ and $\mu_B$ when solving Eqs. \eqref{strneu}. 

\begin{figure*}[t]
    \centering
    \begin{tabular}{c c c}
    \includegraphics[width=0.33\textwidth]{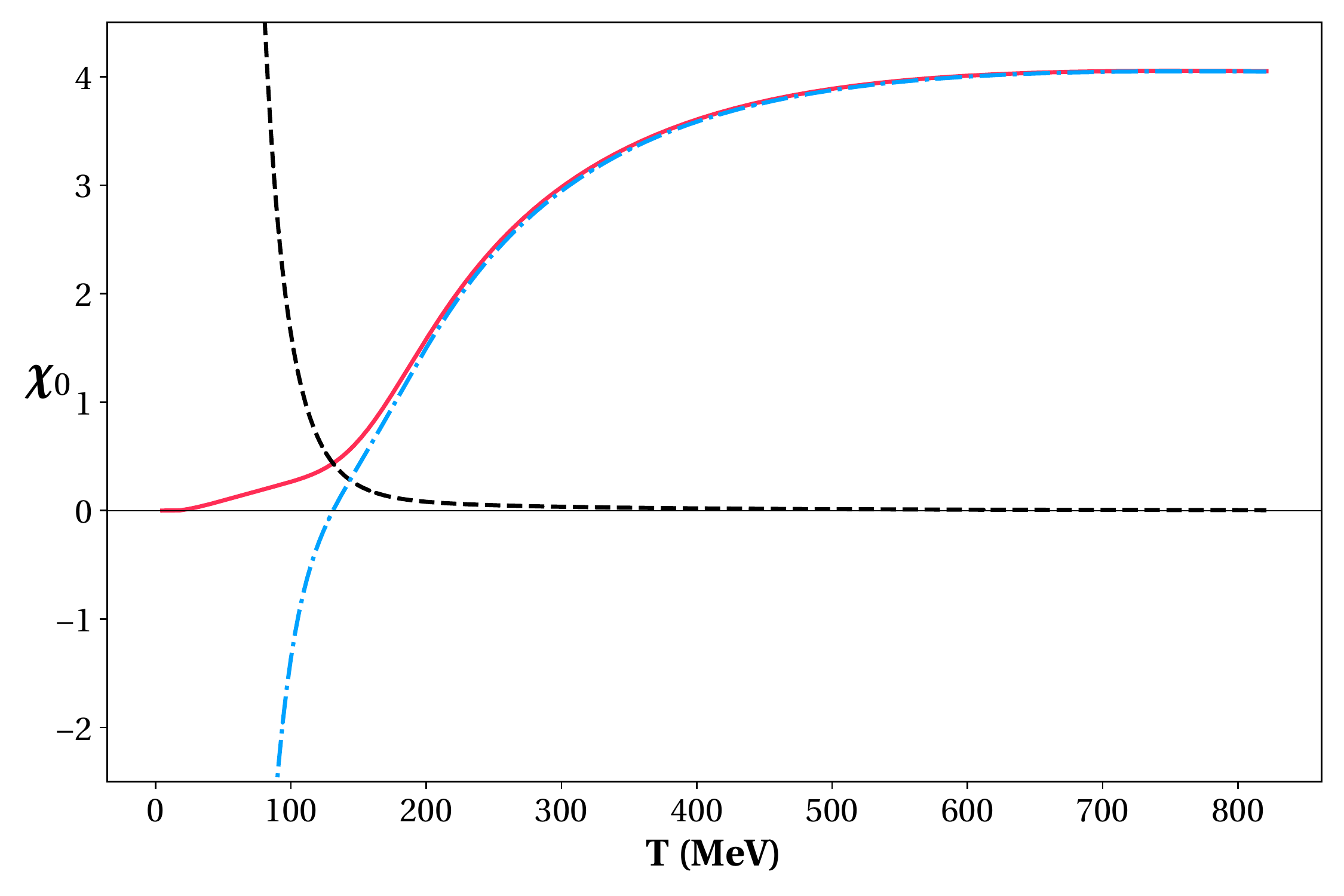} &
    \includegraphics[width=0.33\textwidth]{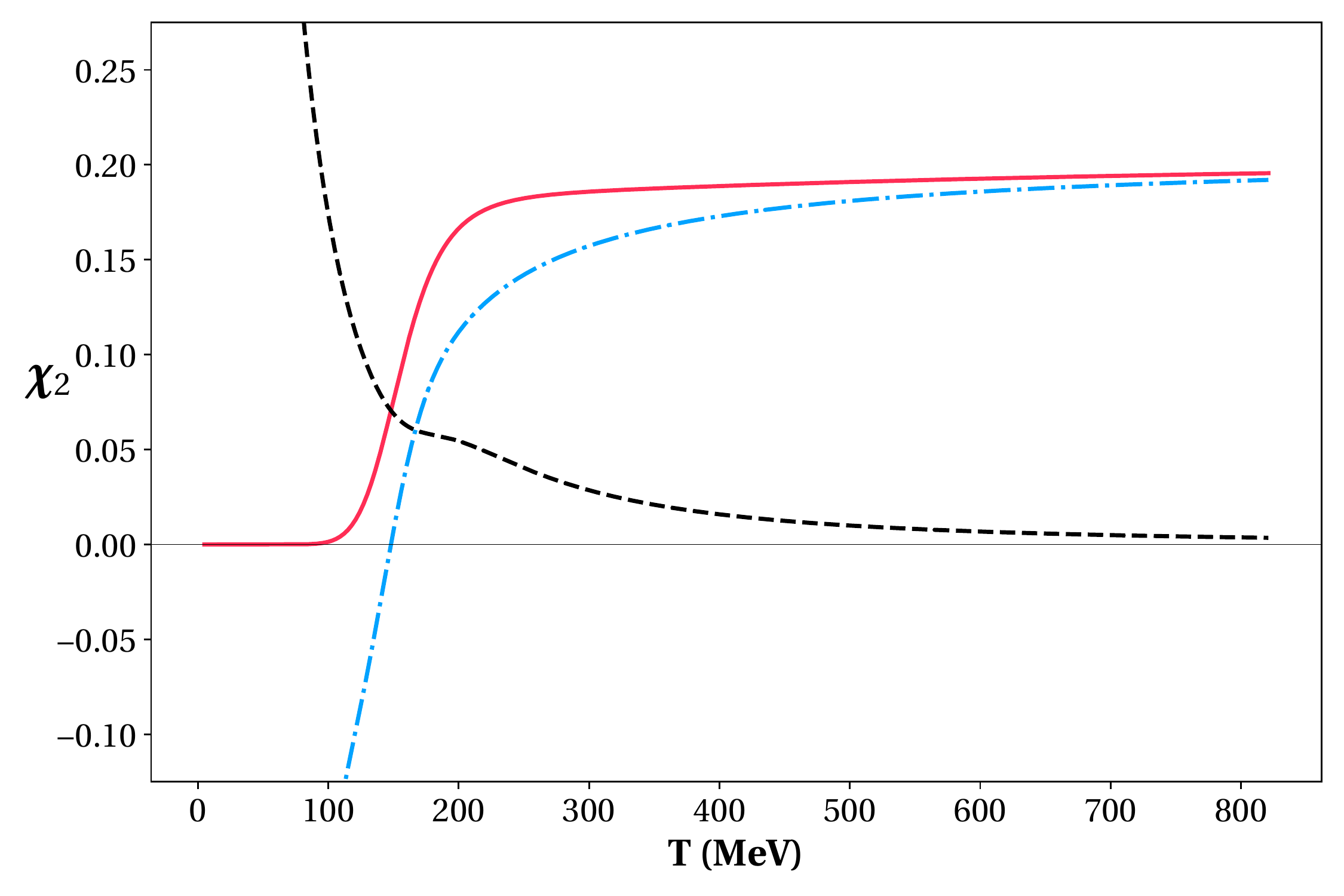} &
    \includegraphics[width=0.33\textwidth]{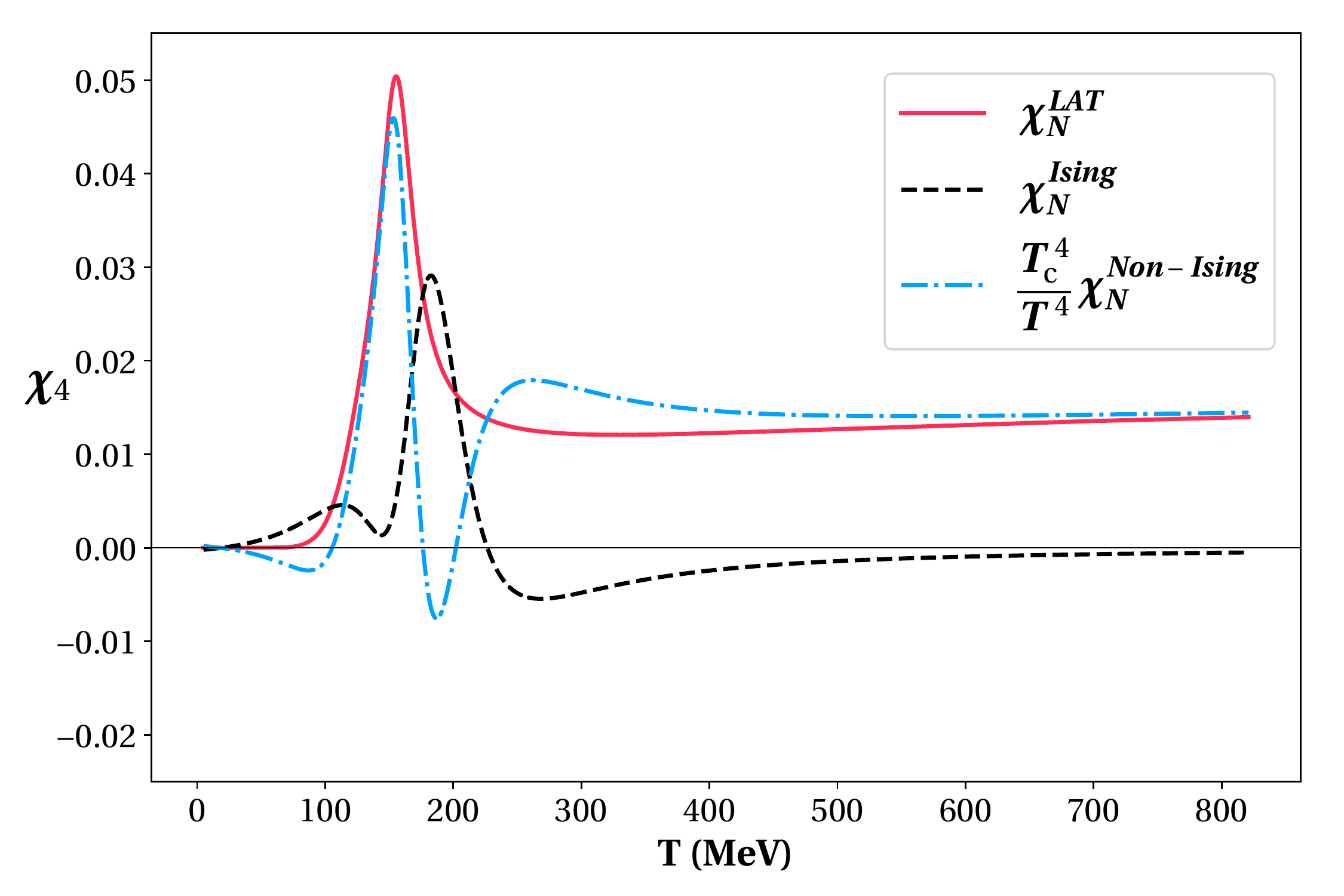}
    \end{tabular}
    \caption{Taylor expansion coefficients of the pressure up to $\mathcal{O}$($\mu_B^4$). The critical contributions from the Ising Model (black, dashed lines) are compared to the lattice QCD results (red, solid lines),  which determine the non-critical terms (blue, dot-dashed lines)  as detailed in Eq. \eqref{coeffmatch}. 
    }
    \label{fig:chis_all}
\end{figure*}

We achieve a continuation of the strangeness-neutral Taylor expansion coefficients from the lattice by merging with the HRG model at T=130 MeV, below which lattice data is unavailable. We additionally require a smooth approach to the Stefan-Boltzmann limit at high temperature. The parametrizations were obtained for $\chi_0$ and $\chi_4$ by employing ratios of polynomials up to the sixth order:
\begin{equation} \label{eq:chi_0_4}
    \begin{split}
    \chi_i(T) &= \frac{a^i_0 + a^i_1/t + a^i_2/t^2 + a^i_3/t^3 + a^i_4/t^4 + a^i_5/t^5 + a^i_6/t^6}{b^i_0 + b^i_1/t + b^i_2/t^2 + b^i_3/t^3 + b^i_4/t^4 + b^i_5/t^5 + b^i_6/t^6},
    \end{split}
\end{equation}
where $t=T/154 \MeV$. For $\chi_2$, a different parametrization was necessary to fit the curve shown in the left panel of Fig. \ref{fig:comp_latt_coeff}:
\begin{equation} \label{eq:chi_2}
    \begin{split}
         \chi_2(T) &= e^{-h_1/t' - h_2/{t'}^2} \cdot f_3 \cdot (1 + \tanh{f_4/t' + f_5}),
    \end{split}
\end{equation}
 where $t'=T/200 \MeV$. The fit coefficients for these functions in terms of inverse temperature are given in Table \ref{tab:fit_params}. In Fig. \ref{fig:chis_all}, we show the parametrizations of each of the Taylor coeffcients for the lattice, Ising, and non-Ising terms up to $\mathcal{O}(\mu_B^4)$. These coefficients are related via Eq. \eqref{coeffmatch}. We show the $\chi_n$'s here for aesthetic reasons, since they are related to the Taylor coefficients via factorials: $c_n = \frac{1}{n!} \chi_n$.

\begin{table*}
\begin{tabular}{c c  c  c  c  c  c  c c}
\hline
\hline
 & & $a_0$ &$a_1$ &$a_2$ &$a_3$ &$a_4$ &$a_5$ &$a_6$ \\
\hline
  & $\chi_0$ & 7.53891 & -6.18858 & -5.37961 & 7.08750 & -0.977970 & 0.0302636 & - \\

  & $\chi_4$ & -390440 & 1131702 & -1219370 & 599799 & -140735 & 14672 & -524 \\
\hline
\hline
 & & $b_0$ &$b_1$ &$b_2$ &$b_3$ &$b_4$ &$b_5$ &$b_6$ \\
\hline
  & $\chi_0$ & 2.24530 & -6.02568 & 15.3737 & -19.6331 & 10.2400 & 0.799479 & - \\

  & $\chi_4$ & -22232566 & 31888765 & 28905535 & -42443011 & -41348102 & 67900917 & -22769197 \\
\hline
\hline
 & & $h_1$ &$h_2$ &$f_3$ &$f_4$ &$f_5$ & & \\
\hline
  & $\chi_2$ & 0.281713 & -0.214339 & 0.103486 & -3.5512 & 4.45949 &  &  \\
\hline
\hline
\end{tabular}
\caption{Parameters for the curves to fit the lattice Taylor coefficients with functional forms given in Eqs. (\ref{eq:chi_0_4},~\ref{eq:chi_2}).}
\label{tab:fit_params}
\end{table*}

We note that the pressure, i.e. $\chi_0$, is the same in this case as in Ref. \cite{Parotto:2018pwx}, since $P(T, \mu_B=0)$ is not altered by the conditions applied for strangeness neutrality. 

As previously stated, this EoS is also compatible with the SMASH transport code \cite{Petersen:2018jag}. This was achieved by utilizing the same hadronic list as the one used in SMASH simulations.  The hadronic list from SMASH contains a total of 400 particles, antiparticles and resonances and includes complete isospin symmetry. In the original formulation of this EoS, the PDG2016+ list, established in Ref. \cite{Alba:2017mqu} as the hadronic list that best matches the partial pressures from Lattice QCD, was utilized. The hadronic degrees of freedom included in SMASH are much fewer than the 739 resonant states in PDG2016+, in which the additional states mainly come from the strange sector. We note that since the HRG model is only used at  temperatures below T=130 MeV we do not see a large effect from the choice of particle list. Since the particle species are the same across the different platforms, this ensures consistency of the thermodynamics throughout the modeling of different stages of heavy-ion collisions. 

\section{Correlation length}\label{sec:corr_leng}

A further update to the BES-EoS is the calculation of the critical correlation length. The search for the critical point in the QCD phase diagram is intrinsically tied to the behavior of the correlation length $\xi$. The divergence of $\xi$ at the critical point is the reason for any non-analyticity of the thermodynamic quantities that make up the EoS. In addition, the correlation length is an important input for hydrodynamic simulations of heavy-ion-collision systems. For instance, in Ref.~\cite{Monnai:2016kud,Dore:2020jye} it was used to describe the critical scaling of the bulk viscosity.  We provide information about the correlation length as given in the 3D Ising model and note that we only provide the critical contribution at this time as the authors are not aware of a calculation of the QCD correlation length on the lattice yet. We adopt the procedure as previously shown in Refs. \cite{Brezin:1976pt, Berdnikov:1999ph, Nonaka:2004pg}, which follows Widom’s scaling form in terms of Ising model variables: 
\begin{equation} \label{corr_length}
    \begin{split}
        \xi^2(r,M) = f^2 |M|^{-2 \nu / \beta} g(x),
    \end{split}
\end{equation}
where $f$ is a constant with the dimension of length, which we set to 1 fm, $\nu$ = 0.63 is the correlation length critical exponent in the 3D Ising Model, $g(x)$ is the scaling function and the scaling parameter is $x$=$\frac{|r|}{|M|^{1/\beta}}$.

In the $\epsilon$-expansion, the function $g(x)$ is given to $\mathcal{O}$($\epsilon^2$) as:
\begin{equation} \label{eq:g_epsilonexp}
    \begin{split}
        g(x) &= g_\epsilon (x) \\
        &= 6^{-2 \nu} z \Big\{ 1 - \frac{\epsilon}{36} [(5 + 6 \ln 3)z - 6(1+z) \ln z ] \\
        &+ \epsilon^2 \Big[\frac{1+2z^2}{72} \ln^2 z + \Big(\frac{z}{18} \Big(z - \frac{1}{2}\Big) (1 - \ln 3) \\
        &- \frac{1}{216} \Big(16z^2 - \frac{47}{3} z - \frac{56}{3} \Big) \Big) \ln z \\
        &+ \frac{1}{216} \Big( \frac{101}{6} + \frac{2}{3}I + 6\ln^2 3 + 4 \ln 3 - 10 \Big) z^2 \\
        &- \frac{1}{216} \Big( 6 \ln^2 3 + \frac{44}{3} \ln 3 + \frac{137}{9} + \frac{8}{3}I \Big) z \Big] \Big\}
    \end{split}
\end{equation}
where $z \equiv \frac{2}{1+x/3}$, $I \equiv \int_0^1 \frac{\ln [x (1-x)]}{1 - x(1-x)}dx \sim -2.344$, and $\epsilon = 4 - d$, where $d=3$ is the spatial dimension.

However, when $x$ becomes large, the above expression cannot be used, and one must use the asymptotic form:
\begin{equation} \label{eq:g_asymptotic}
    \begin{split}
        g(x) = g_{\text{large}} (x) = \Big( \frac{1}{3+x} \Big)^{2\nu}.
    \end{split}
\end{equation}
This requires a smooth merging between the different formulations of $g(x)$, in order to provide a well-behaved result for correlation length. Analogously to what was done in Ref. \cite{Nonaka:2004pg}, we perform this smooth merging around $x=5$. We make use of a hyperbolic tangent at the level of the QCD variables, i.e. after we have already changed from Ising Model variables to QCD ones as described in Section II. Further details on the procedure for the smooth merging can be found in Appendix \ref{sec:smoothmerg}. 
\section{Results}\label{sec:results}
\begin{figure*}
    \centering
    \includegraphics[width=0.48\textwidth]{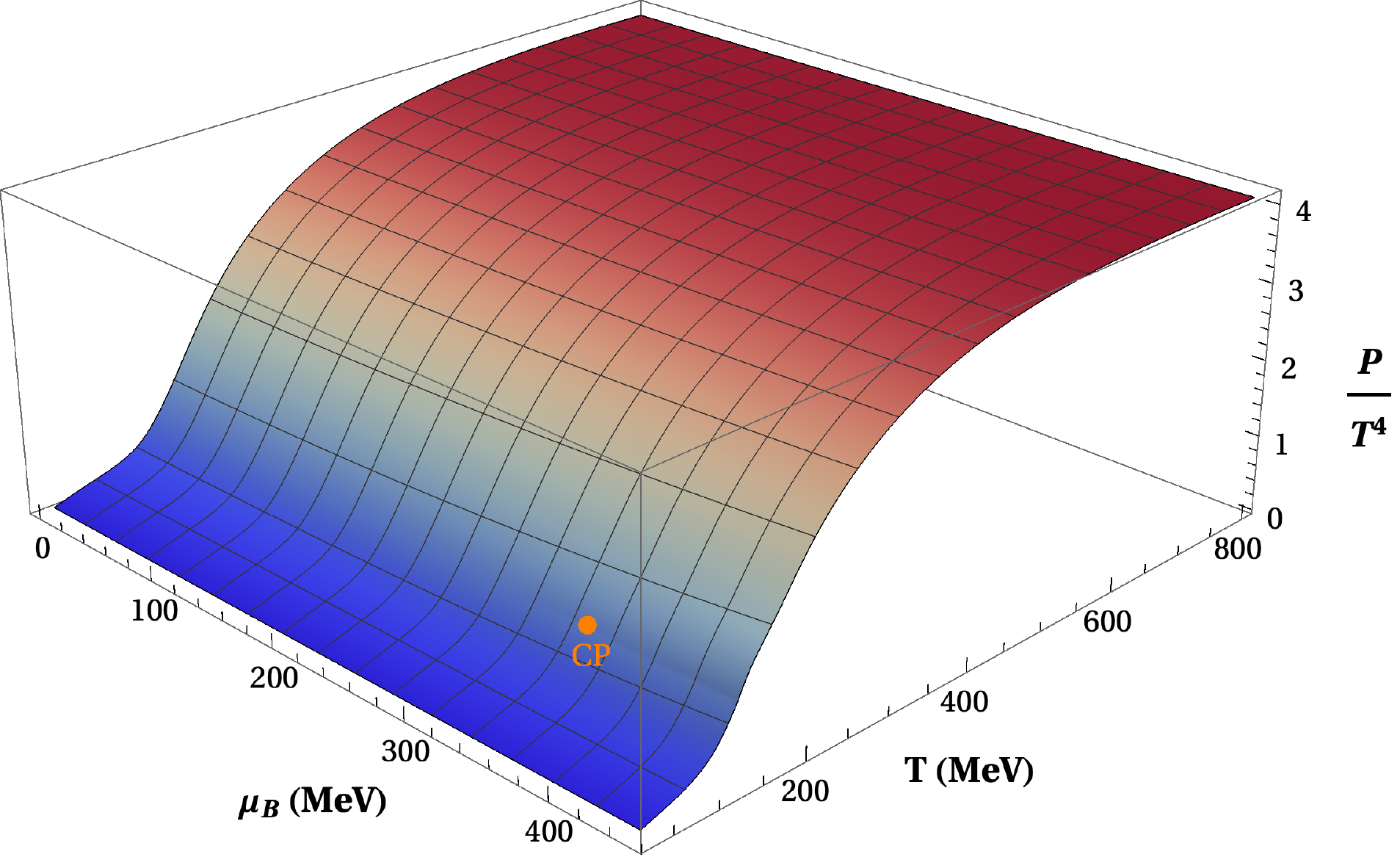}
    \includegraphics[width=0.5\textwidth]{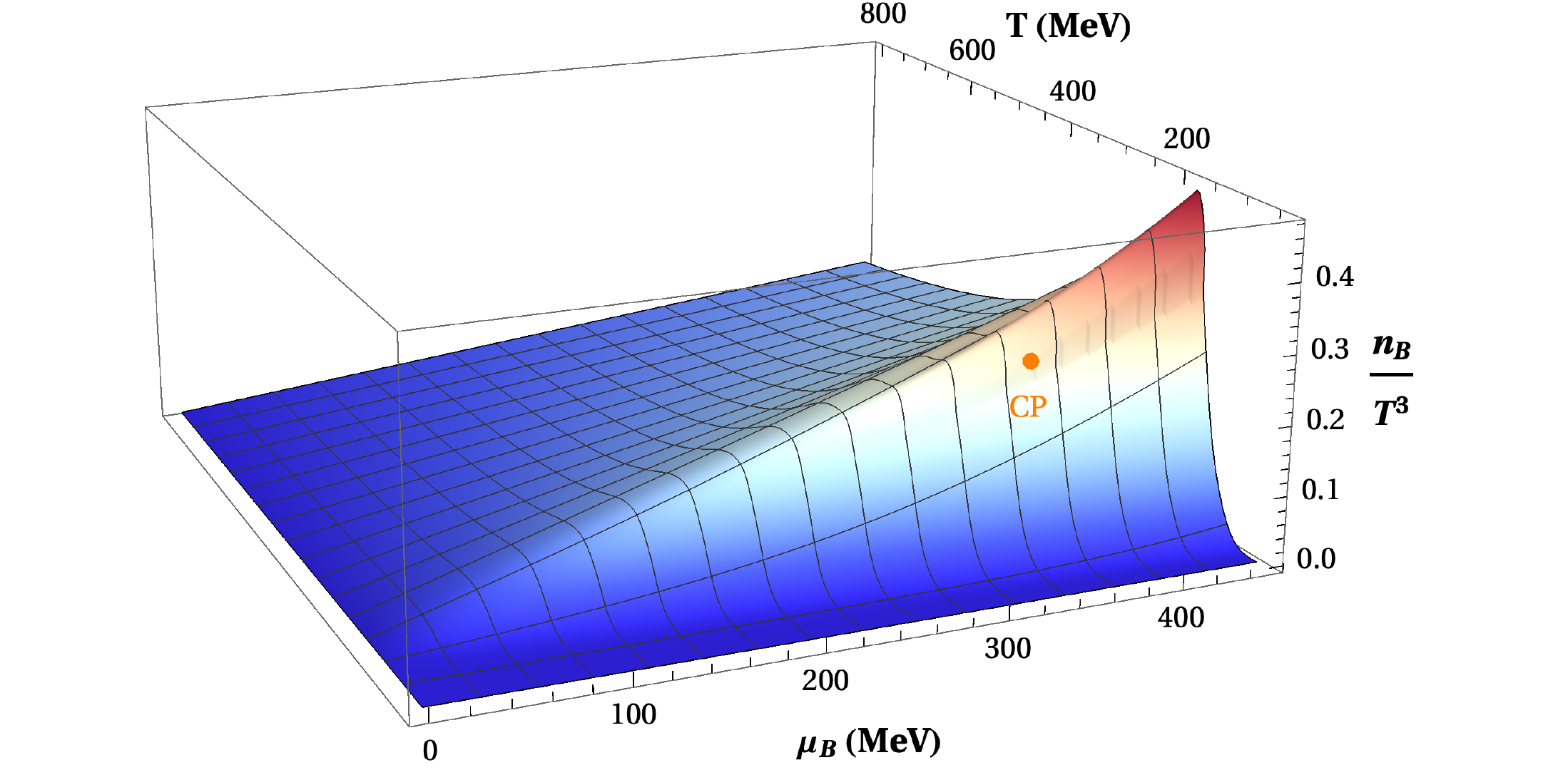}
    \caption{Left: The full QCD pressure for the choice of parameters consistent with Ref. \cite{Parotto:2018pwx}, listed in Section \ref{sec:scal_EoS}.  Right: The baryon density for the same choice of parameters.  }
    \label{fig:press}
\end{figure*}

\begin{figure*}
    \centering
    \includegraphics[width=0.49\textwidth]{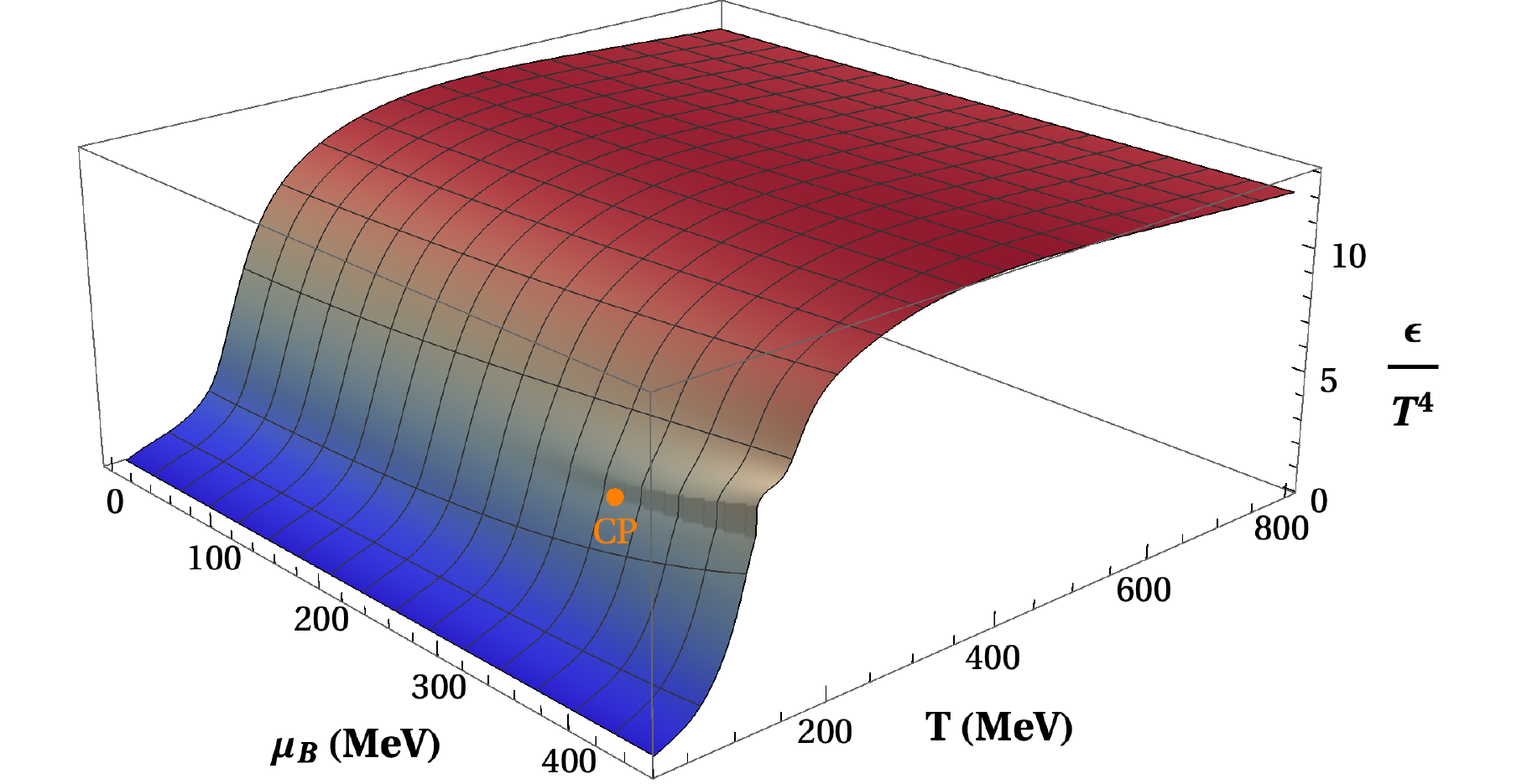}
    \includegraphics[width=0.49\textwidth]{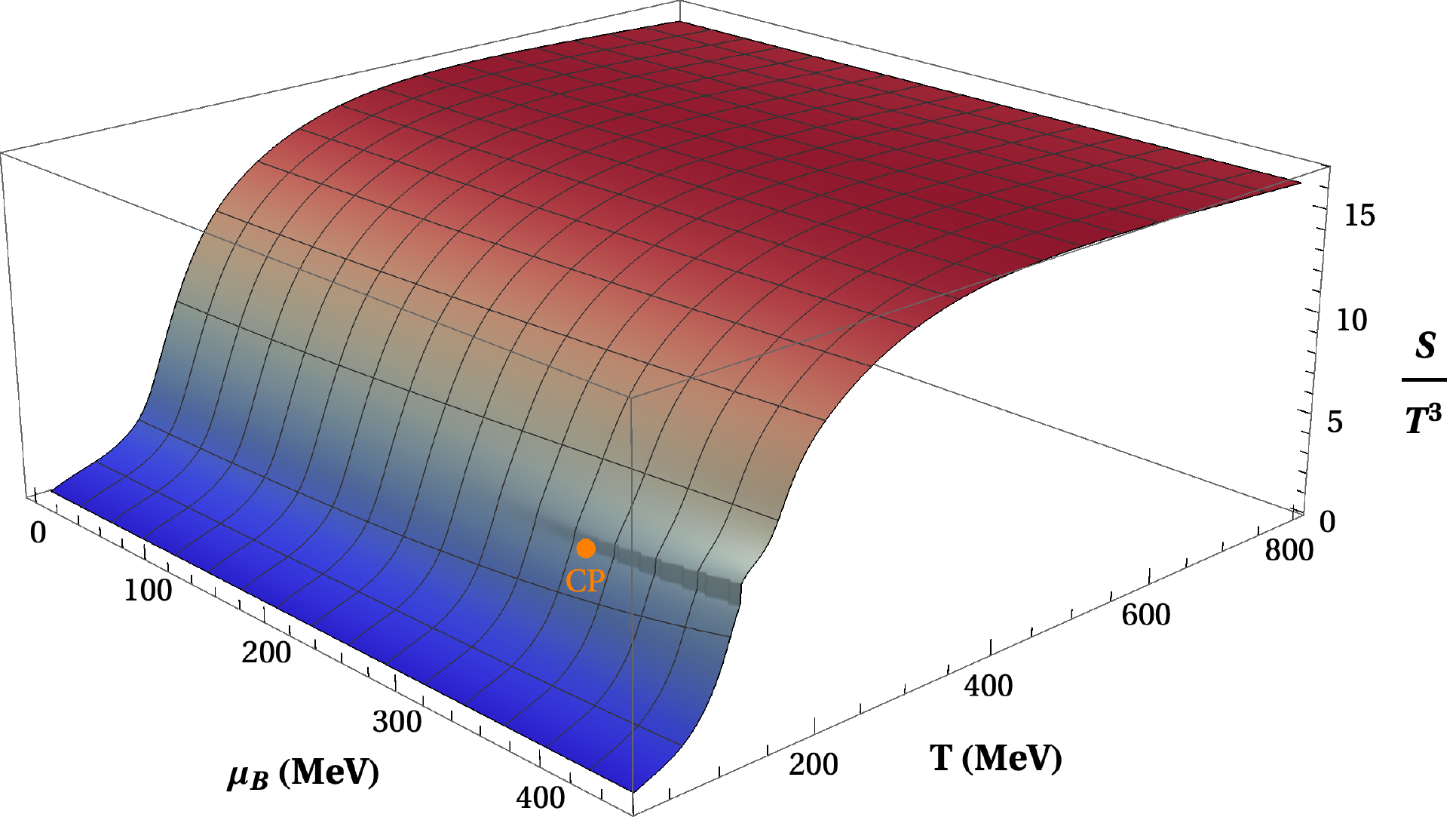}
    \caption{Left: The energy density for the choice of parameters consistent with Ref. \cite{Parotto:2018pwx}, listed in Section \ref{sec:scal_EoS}. Right: The entropy density for the same choice of parameters.}
    \label{fig:enerdens}
\end{figure*}

\begin{figure*}
    \centering
    \includegraphics[width=0.49\textwidth]{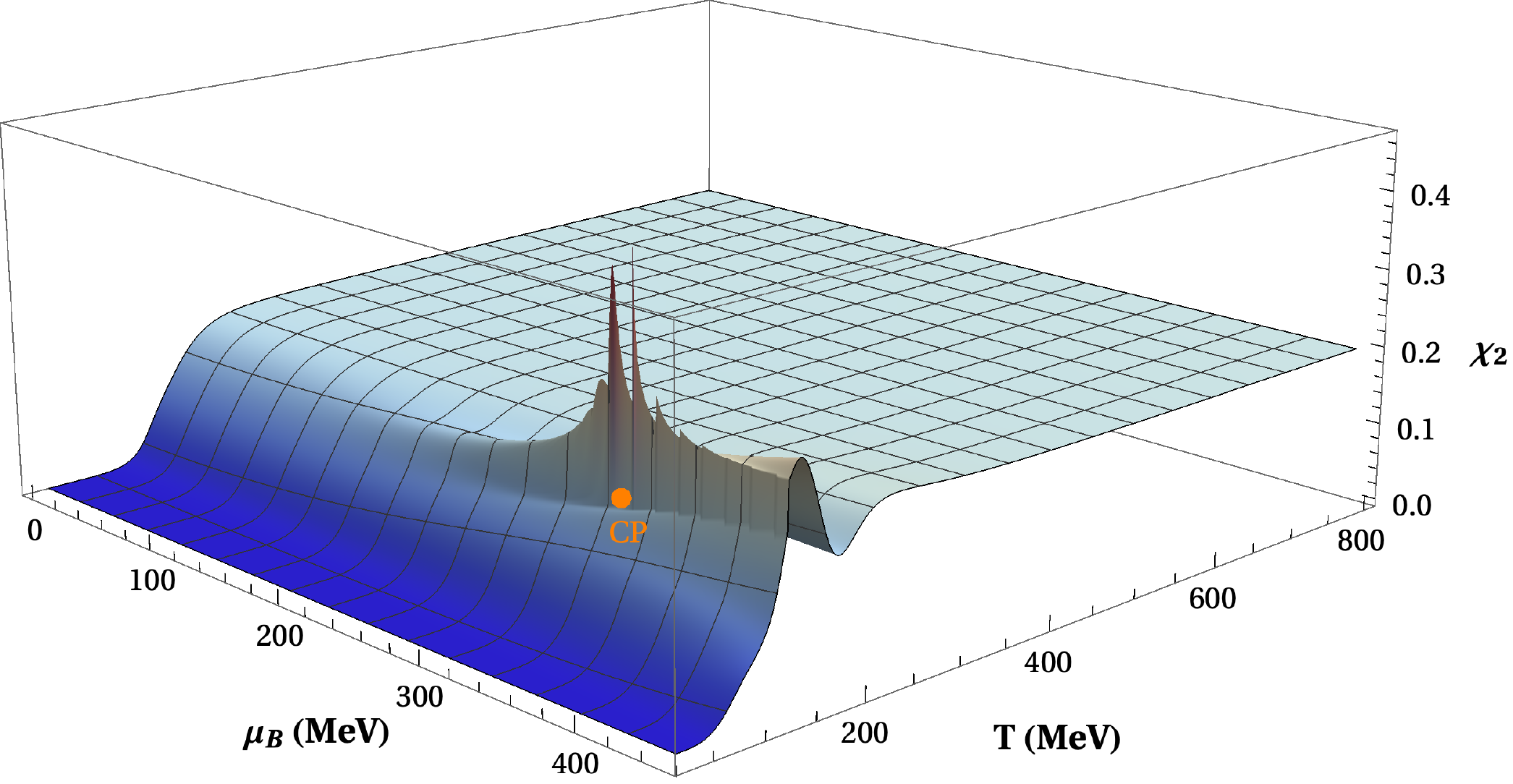}
    \includegraphics[width=0.49\textwidth]{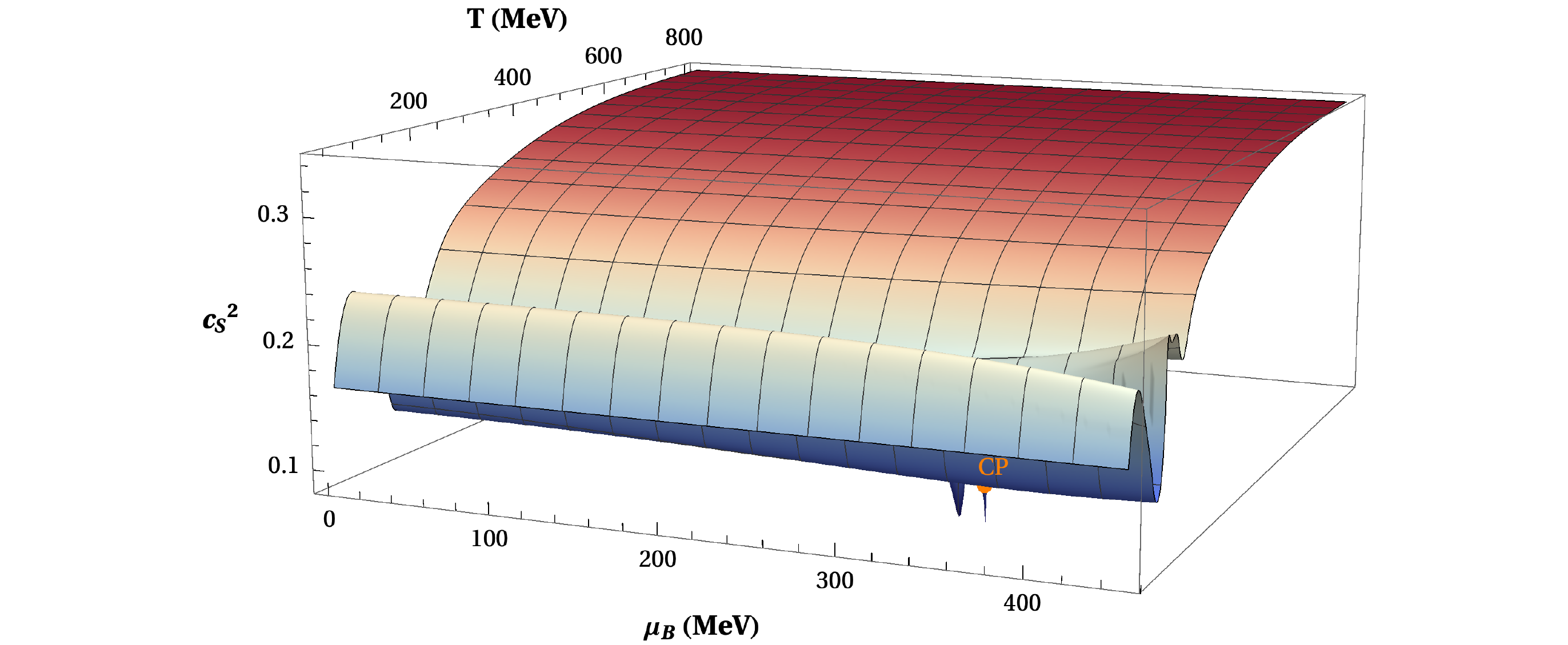}
    \caption{Left: The 2nd baryon susceptibility for the choice of parameters consistent with Ref. \cite{Parotto:2018pwx}, listed in Section \ref{sec:scal_EoS}. Right: The speed of sound for the same choice of parameters.}
    \label{fig:spsound}
\end{figure*}

We begin by showing all thermodynamic quantities calculated within this framework. The Taylor-expanded pressure is shown in the left panel of Fig. \ref{fig:press}, while the derivatives of the pressure (see Eq. \eqref{eq:thermodef} for definitions) are shown in the subsequent plots. In particular, the right panel of Fig. \ref{fig:press} shows the baryonic density, the two panels of Fig. \ref{fig:enerdens} show the energy density and entropy density, while those of Fig. \ref{fig:spsound} show the second order baryon number susceptibility and the speed of sound, respectively. 
Such quantities can be obtained from the pressure according to the following relationships:
\begin{equation} \label{eq:thermodef}
\centering
    \begin{split}
         \,\,\,\, \frac{n_B}{T^3}&=\frac{1}{T^3} 
        \left( \frac{\partial P}{\partial \mu_B} \right)_T, \\
        \frac{\chi_2^B}{T^2}&=\frac{1}{T^2} 
        \left( \frac{\partial^2 P}{\partial \mu_B^2} \right)_T, \\
        \frac{S}{T^3}&=\frac{1}{T^3} 
        \left( \frac{\partial P}{\partial T} \right)_{\mu_B}, \\
        \frac{\epsilon}{T^4}&=\frac{S}{T^3} - \frac{P}{T^4} + \frac{\mu_B}{T} \frac{n_B}{T^3}, \\
        c_S^2 &= \left( \frac{\partial P}{\partial \epsilon} \right)_{S/n_B}.
    \end{split}
\end{equation}

The location of the critical point is indicated on each of these graphs to guide the reader to the critical region. As expected, the pressure is a smooth function of \{T,  $\mu_B$\} in the crossover region, while it shows a slight kink for chemical potentials larger the critical point one. The derivatives of the pressure help to reveal the features of the critical region, with an enhancement of criticality with increasing order of derivatives. For example, consider the baryon density shown in the right panel of Fig. \ref{fig:press}, compared to the second cumulant of the baryon number shown in the left panel of Fig. \ref{fig:spsound}: the former exhibits a discontinuity due to the presence of the critical point while the latter is shown to diverge at the critical point, as expected. This intensification of critical features is due to the dependence on increasing powers of the correlation length, $\xi$, of higher order cumulants.


\begin{figure*}
    \centering
    \includegraphics[width=0.5\textwidth]{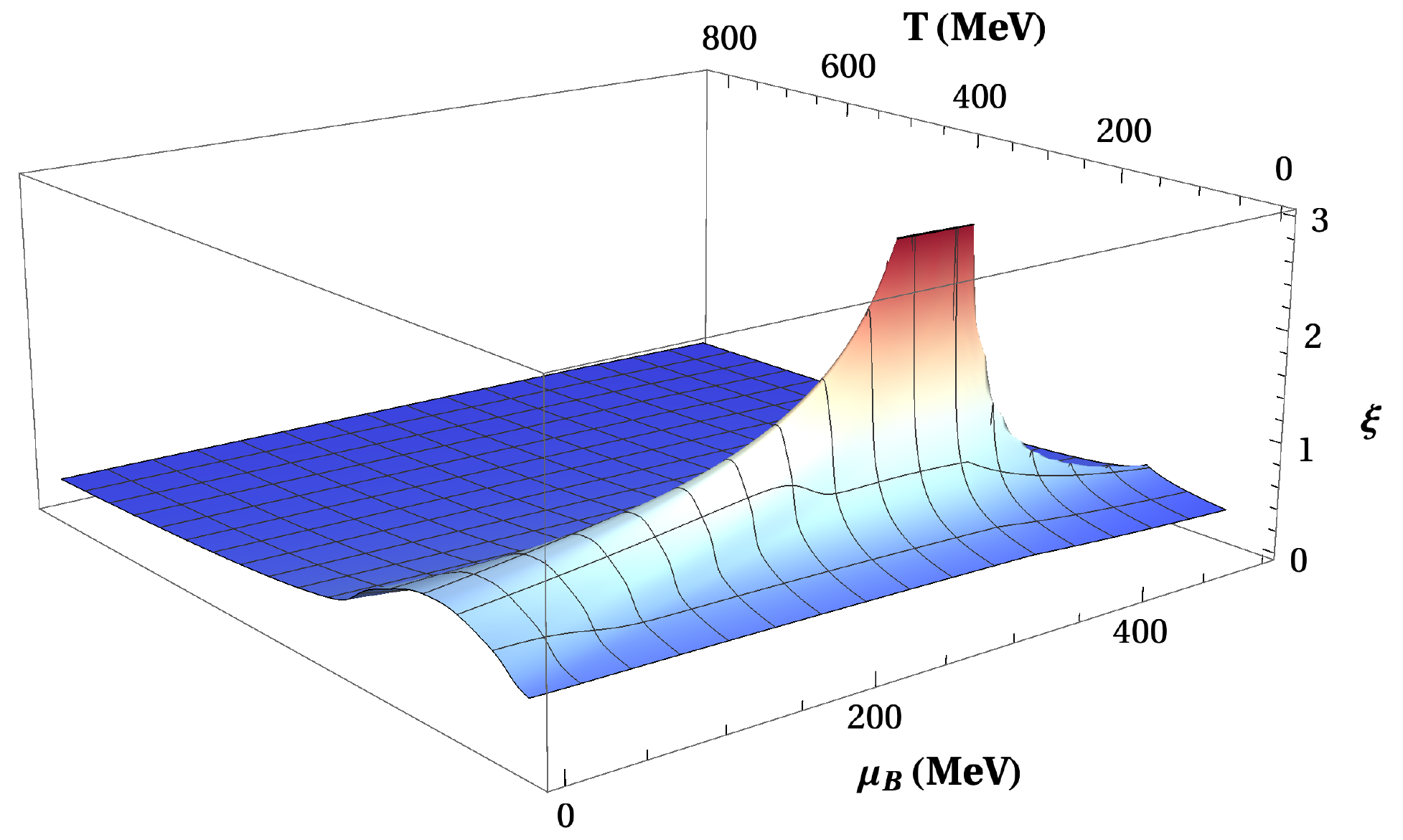}
    \caption{The critical contribution to the correlation length for the choice of parameters consistent with Ref. \cite{Parotto:2018pwx}, listed in Section \ref{sec:scal_EoS}.}
    \label{fig:corrleng}
\end{figure*}

We show next in Fig. \ref{fig:corrleng} the critical contribution to the correlation length, calculated as described in section IV, as a function of \{$T$,  $\mu_B$\}.
The smoothly merged correlation length is uniform everywhere in the phase diagram, except in the critical region. There, the correlation length increases and then diverges at the critical point itself, following the expected scaling behavior.

\begin{figure*}
    \centering
    \includegraphics[width=0.49\textwidth]{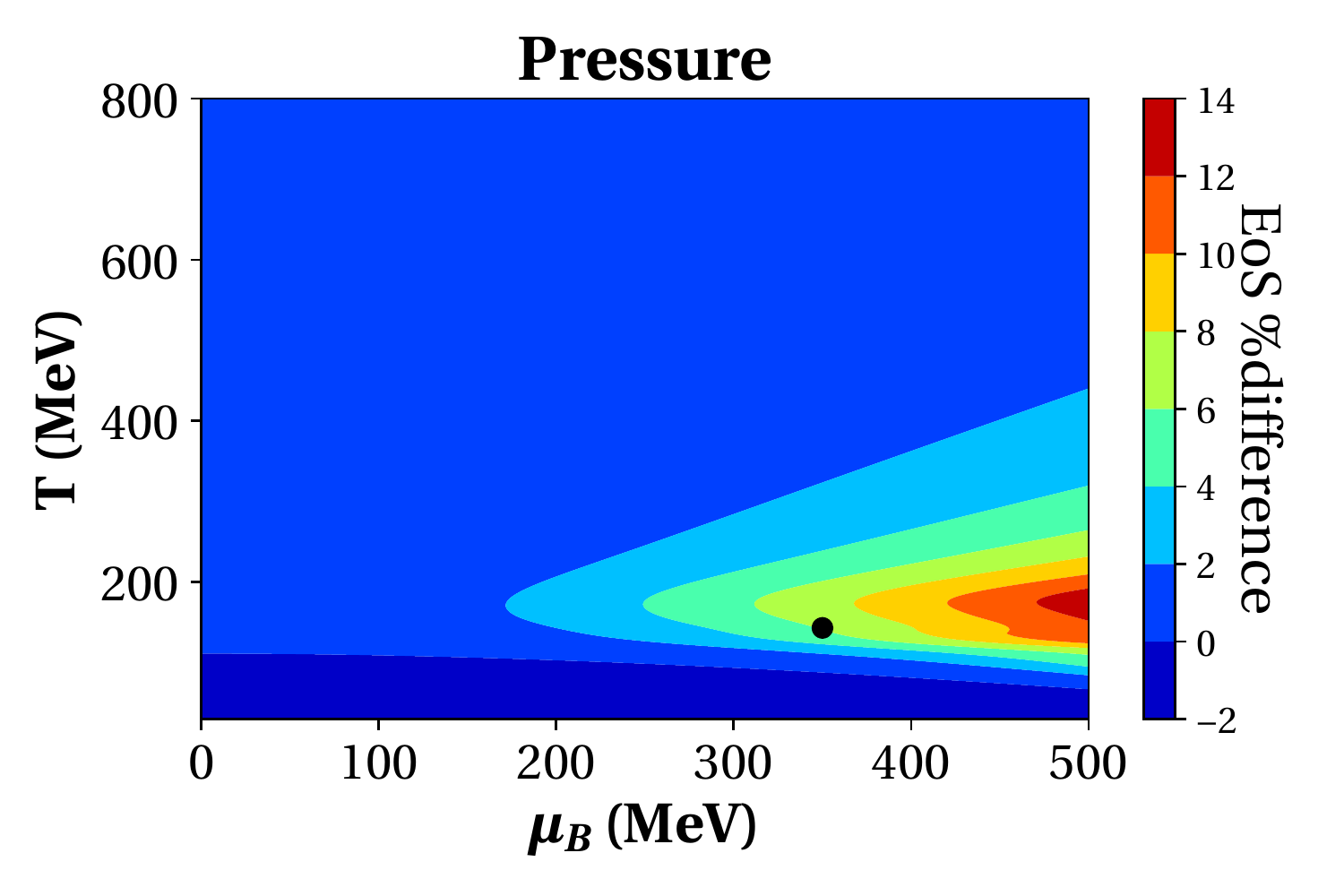}
    \includegraphics[width=0.49\textwidth]{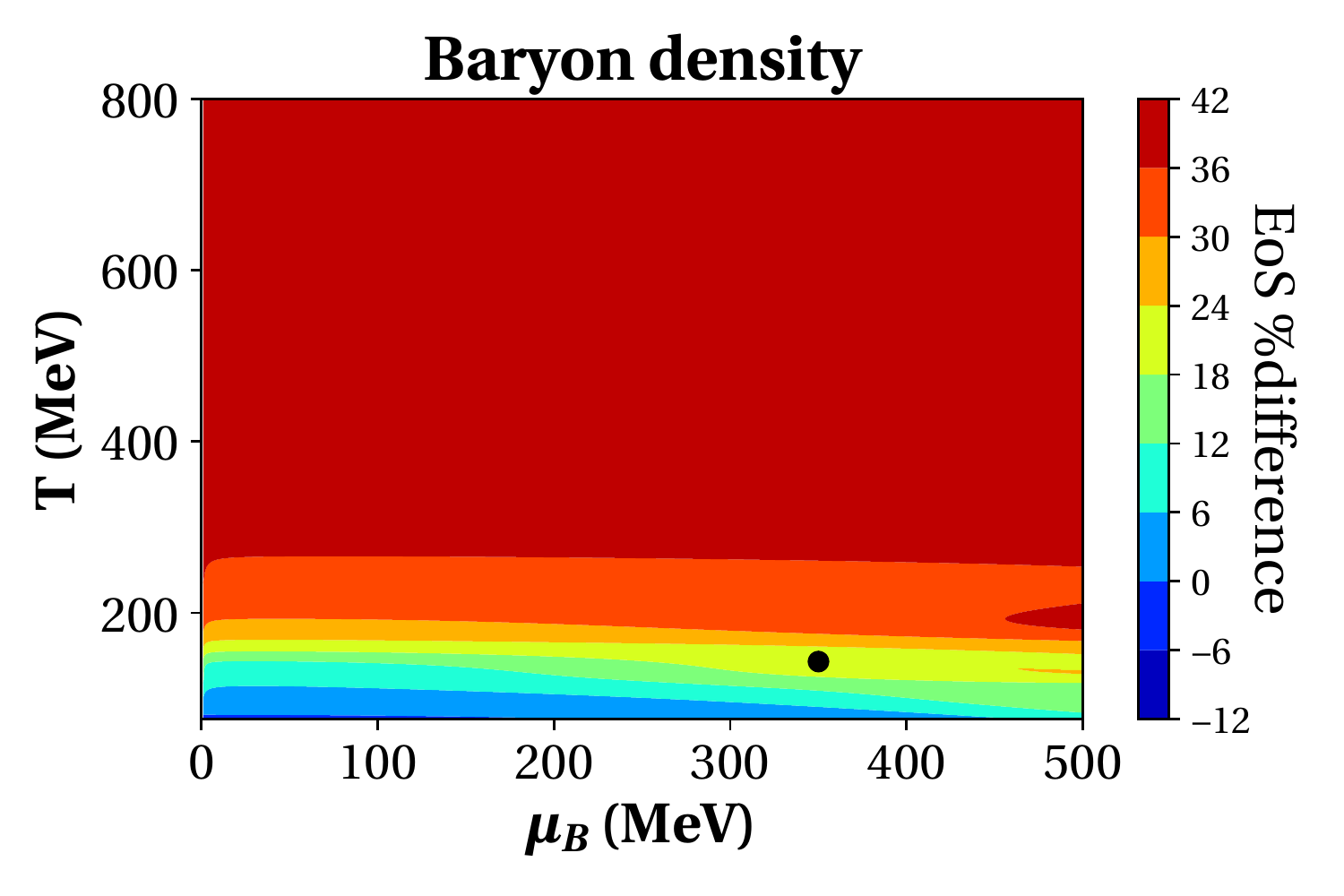}
    \caption{Left: Relative percent difference for the pressure between the original EoS formulation (Ref. \cite{Parotto:2018pwx}) and the updated strangeness-neutral version presented in this manuscript. Right: Relative percent difference for the baryon density between the original EoS formulation (Ref. \cite{Parotto:2018pwx}) and the updated strangeness-neutral version presented in this manuscript. In both plots, the point marks the location of the critical point as listed in Sec. \ref{sec:scal_EoS}. For more details, see text.}
    \label{fig:pressdiff}
\end{figure*}

\begin{figure*}
    \centering
    \includegraphics[width=0.49\textwidth]{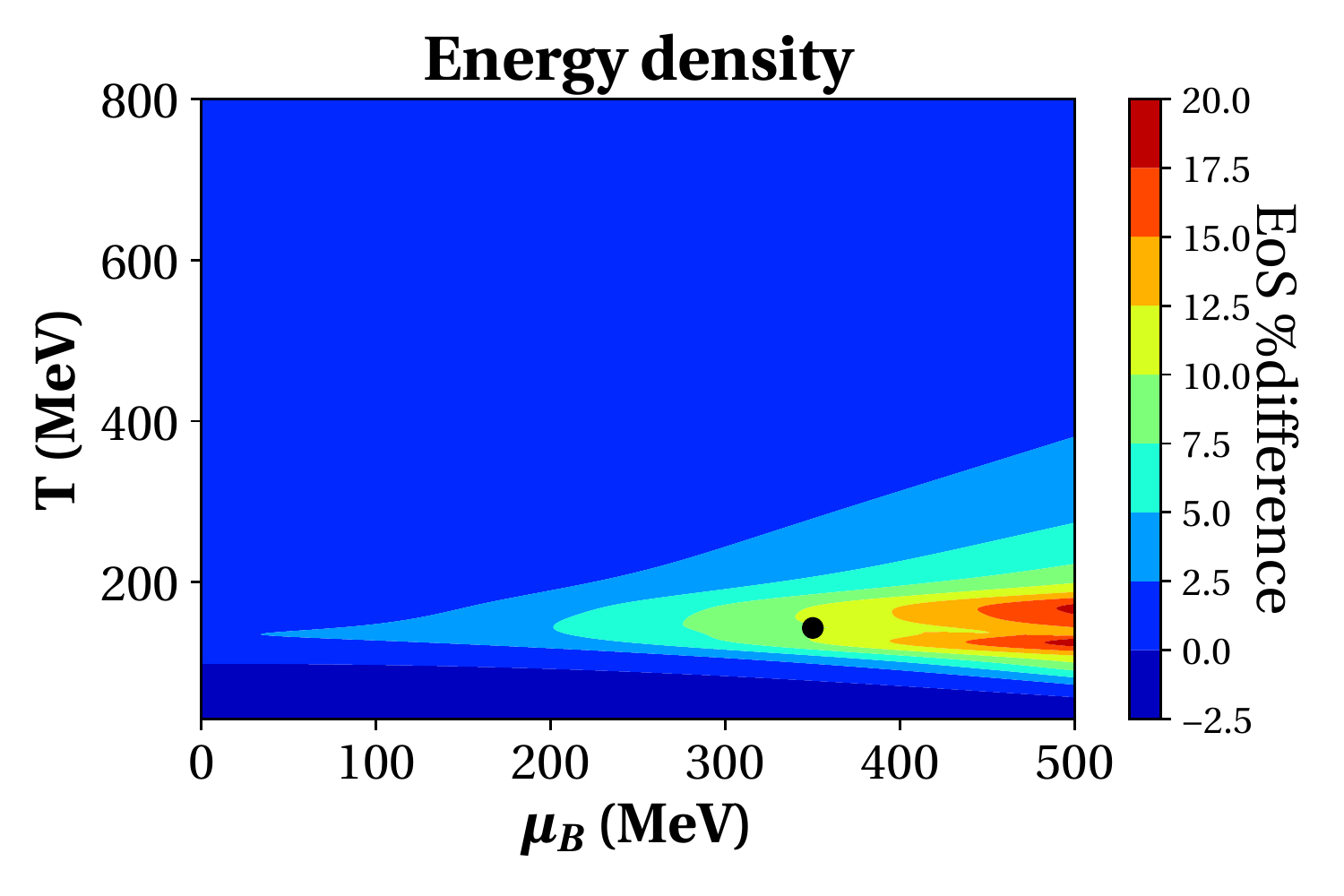}
    \includegraphics[width=0.49\textwidth]{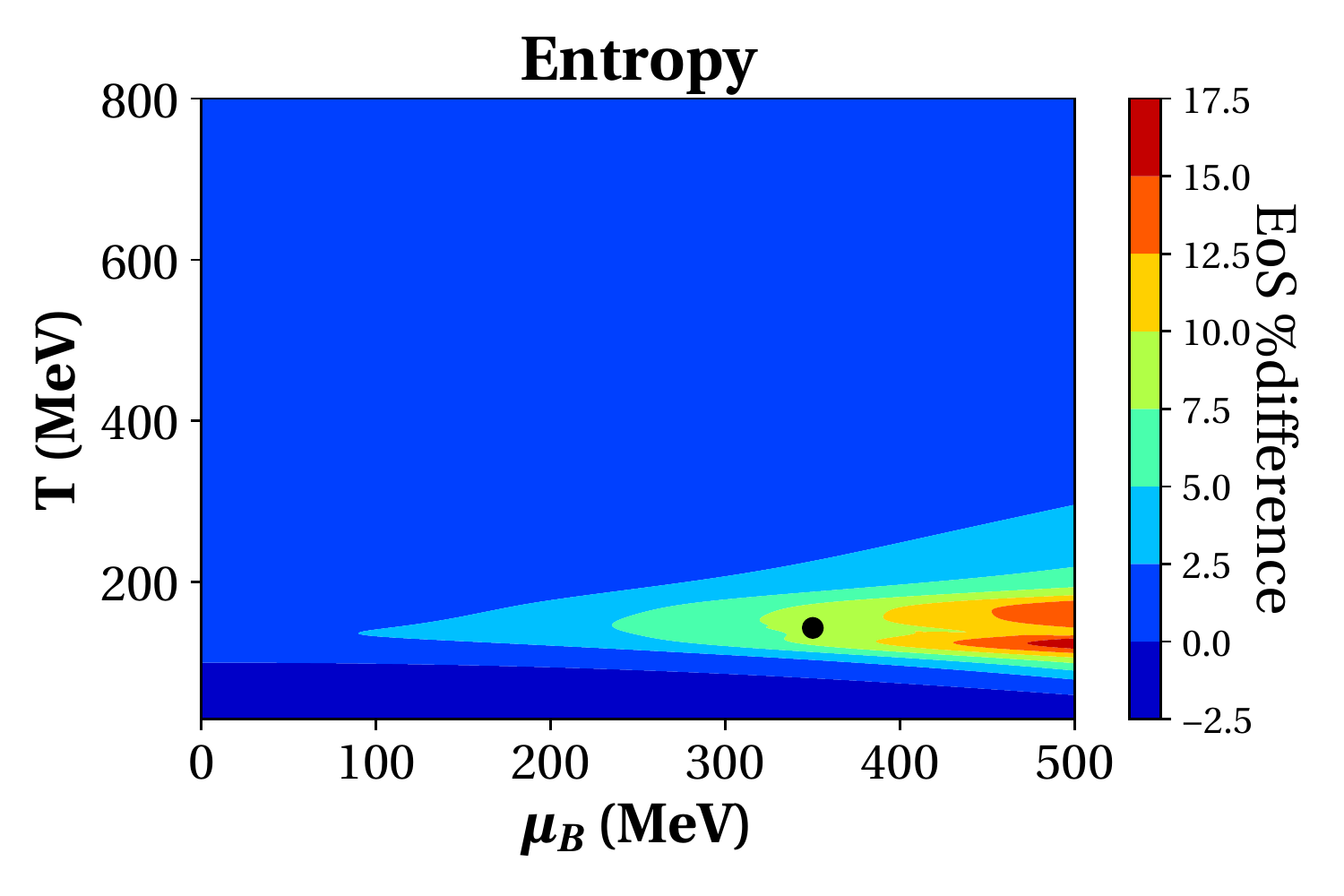}
    \caption{Left: Relative percent difference for the energy density between the original EoS formulation (Ref. \cite{Parotto:2018pwx}) and the updated strangeness-neutral version presented in this manuscript. Right: Relative percent difference for the entropy between the original EoS formulation (Ref. \cite{Parotto:2018pwx}) and the updated strangeness-neutral version presented in this manuscript. In both plots, the point marks the location of the critical point as listed in Sec. \ref{sec:scal_EoS}. For more details, see text.}
    \label{fig:enerdensdiff}
\end{figure*}

\begin{figure*}
    \centering
    \includegraphics[width=0.49\textwidth]{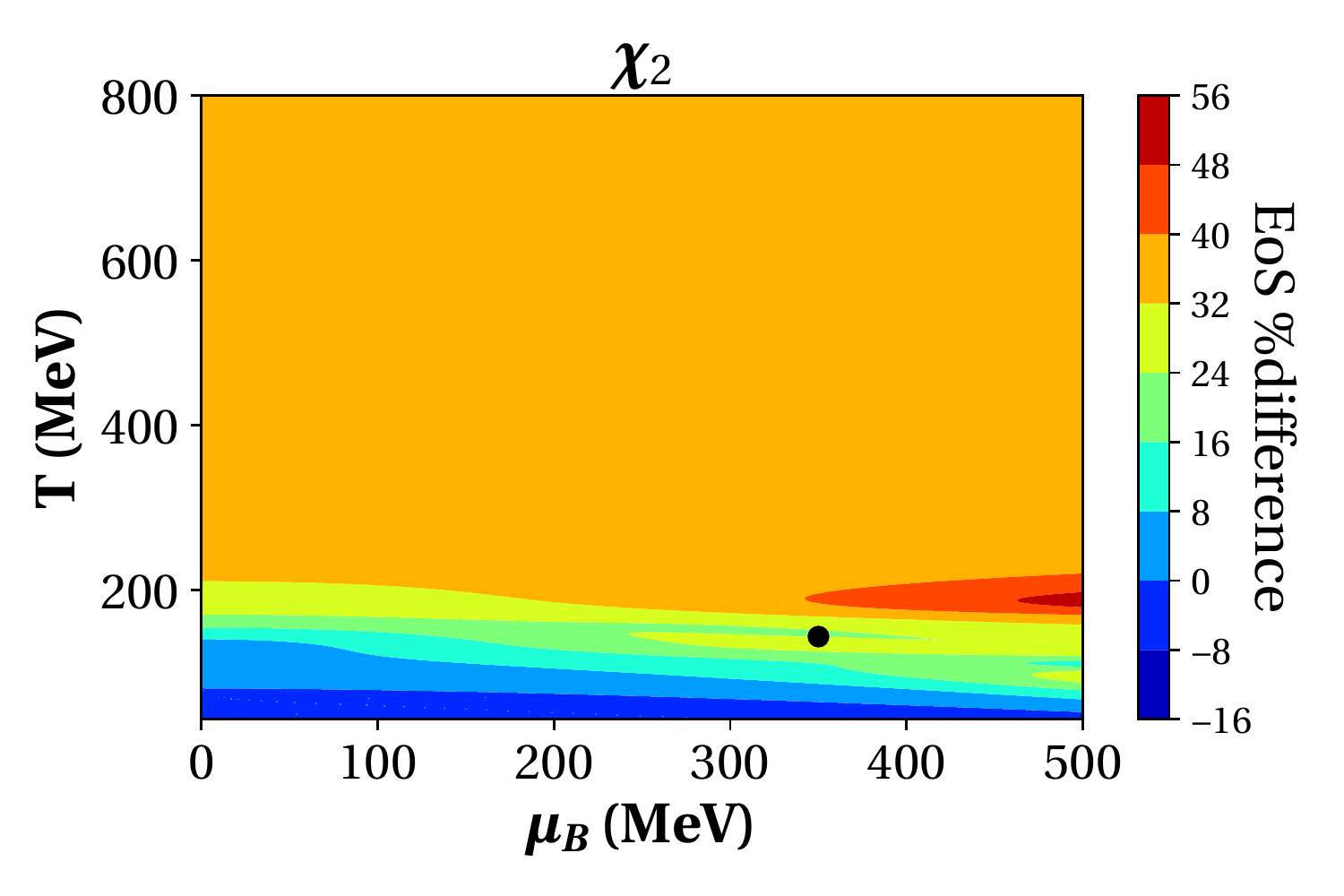}
    \includegraphics[width=0.49\textwidth]{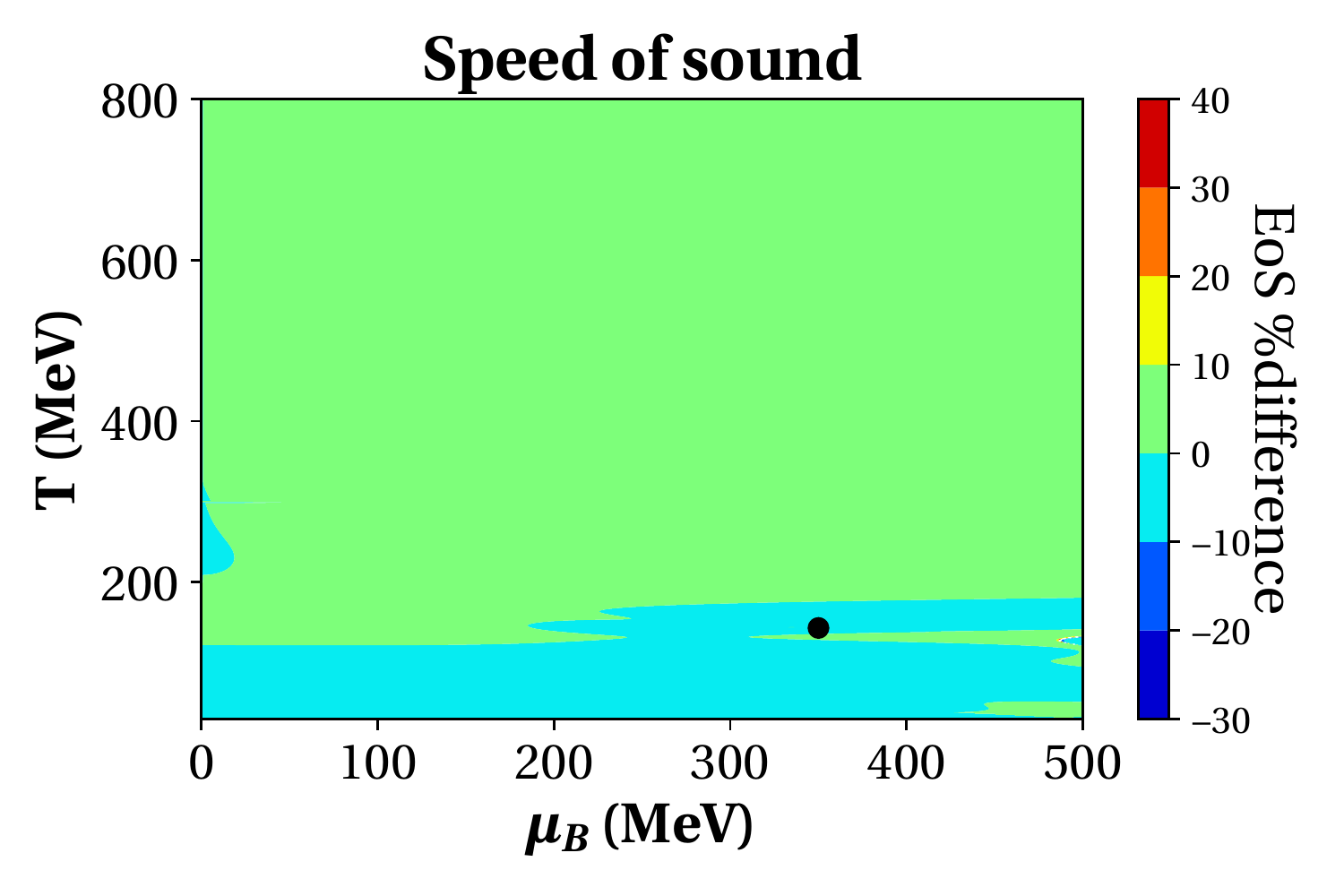}
    \caption{Left: Relative percent difference for the second baryon susceptibility between the original EoS formulation (Ref. \cite{Parotto:2018pwx}) and the updated strangeness-neutral version presented in this manuscript. Right: Relative percent difference for speed of sound between the original EoS formulation (Ref. \cite{Parotto:2018pwx}) and the updated strangeness-neutral version presented in this manuscript. In both plots, the point marks the location of the critical point as listed in Sec. \ref{sec:scal_EoS}. For more details, see text.}
    \label{fig:spssoundiff}
\end{figure*}

We provide a quantitative assessment of the differences between the original EoS with $\mu_S=\mu_Q=0$ and the new system with strangeness neutrality conditions. Contour plots comparing each of the thermodynamic outputs of the code are shown in Figs. 
\ref{fig:pressdiff} - \ref{fig:spssoundiff}.

In these plots, we show the percent difference between the results from the previous formulation and the new one. In general, the largest deviations are found at high baryonic chemical potential, which is to be expected as the values of $\mu_S$ and $\mu_Q$ are monotonically increasing functions of $\mu_B$ (and therefore their value at high $\mu_B$ is the farthest away from 0).  
For the baryon density and the second baryon number susceptibility, we note that the large difference at high temperatures is due to the different high temperature limits of these quantities in the case of $\mu_S=\mu_Q=0$ and strangeness neutrality, as shown for $\chi_2$ in Fig. \ref{fig:comp_latt_coeff}. The negative differences at low temperature can be attributed to the fact that the SMASH treatment assumes complete isospin symmetry. For example, the neutral pion, $\pi^0$, appears with a mass of 138 MeV, while Ref. \cite{Parotto:2018pwx} uses the particle list called PDG16+ (introduced in Ref. \cite{Alba:2017mqu}), which includes the physical mass for the neutral pion: $m_{\pi^0} = 135$ MeV. For details, see Ref. \cite{Weil:2016zrk}.

\begin{figure*}
    \centering
    \begin{tabular}{c c}
    \includegraphics[width=0.5\textwidth]{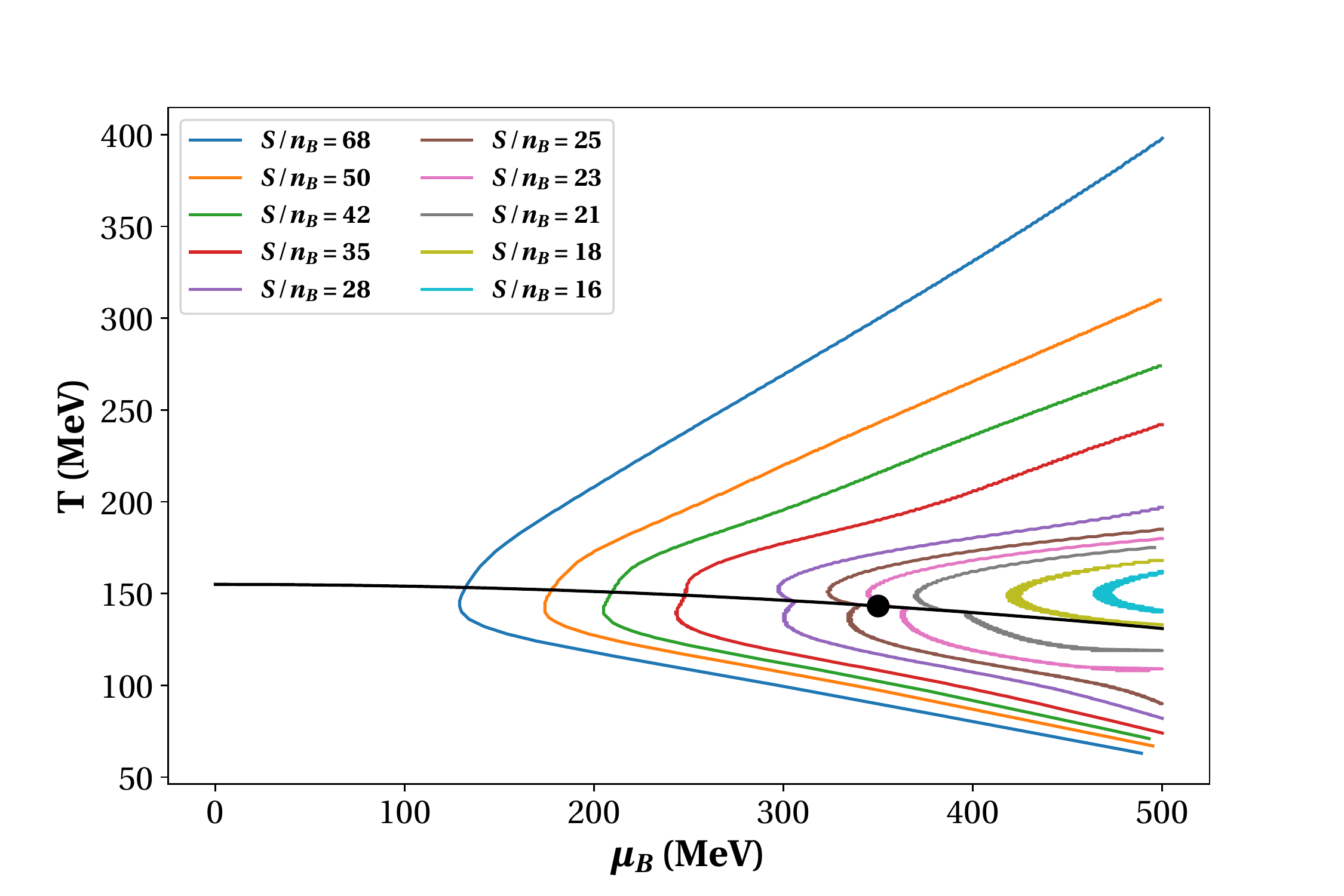} &
    \includegraphics[width=0.5\textwidth]{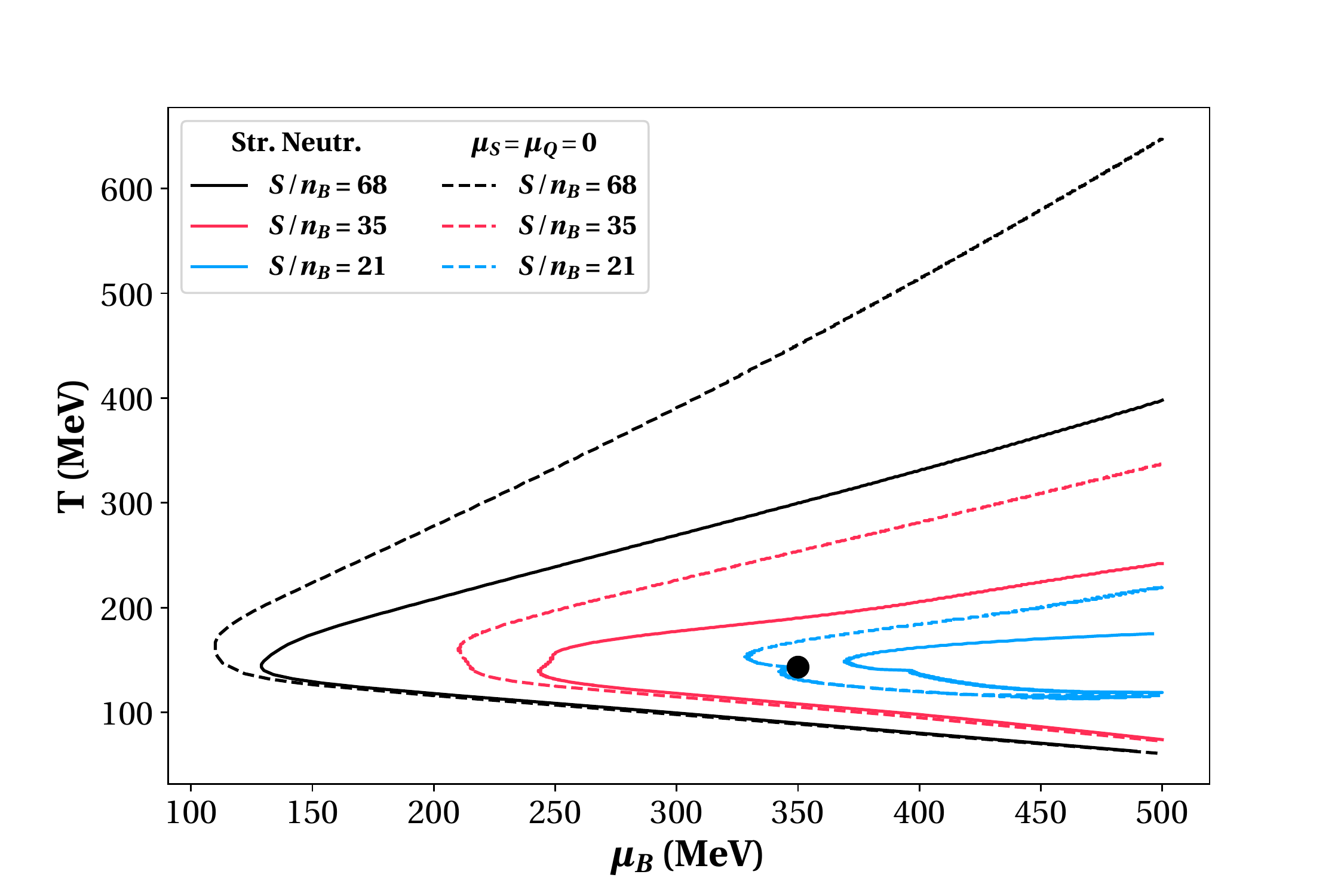}
    \end{tabular}
    \caption{Isentropic trajectories in the QCD phase diagram. Left: Lines of constant $S/n_B$ with the transition line shown in black. Right: Comparison of the isentropes in the case of strangeness neutrality (dashed lines) and in the original formulation with $\mu_S=\mu_Q=0$ from Ref. \cite{Parotto:2018pwx} (solid lines). In both plots, the point marks the location of the critical point as listed in Sec. \ref{sec:scal_EoS}.}
    \label{fig:isen}
\end{figure*}

In Fig. \ref{fig:isen}, we present the isentropic trajectories in the QCD phase diagram, which represent lines of constant $S/n_B$. If the isentrope passes through the critical region, it exhibits a disturbance where there is a jump in both the entropy and the baryon density. Fig.~\ref{fig:isen} also shows the comparison between the new isentropes including strangeness neutrality and the previous ones from the literature \cite{Parotto:2018pwx}. While both exhibit features of the critical point, their trajectories through the phase diagram are quite different in the two cases.  Since the isentropes can be understood to represent the path of the heavy-ion system through the phase diagram in the absence of dissipation, we note that this plot in particular shows the importance of incorporating strangeness neutrality into the EoS. Strangeness neutrality pushes the trajectories to larger $\mu_B$ for the same value of $T$, which has implications for the initial conditions of heavy-ion collisions.

\section{Conclusions}\label{sec:concl}

We provide an equation of state with strangeness neutrality and fixed baryon-number-to-electric-charge ratio, realizing the experimental conserved charge conditions, to be used in hydrodynamic simulations of heavy-ion collisions. Not only does this framework probe a slice of the QCD phase diagram covered by the experiments, but also includes exclusively the hadronic states used in the SMASH hadronic afterburner \cite{Petersen:2018jag}. Additionally, we compare this new equation of state with the original version that first matched the Taylor expansion coefficients from Lattice QCD and implemented critical features based on universality arguments \cite{Parotto:2018pwx}, but at $\mu_S=\mu_Q=0$. Furthermore, we incorporate a calculation of the correlation length in the 3D Ising model, which exhibits the expected scaling behavior and could be used to calculate the critical scaling of transport coefficients at the critical point.

\section*{Acknowledgements}
The authors would like to thank Hannah Elfner, Anna Sch\"afer, and Dmytro Oliinychenko for their assistance in obtaining the hadronic list from SMASH. We would also like to thank Volker Koch, Volodymyr Vovchenko, and Travis Dore for fruitful discussions.
Furthermore, we would like to acknowledge our fellow BEST collaboration members for helping motivate this work at the previous collaboration meeting. This material is based upon work supported by the
National Science Foundation under grant no. PHY1654219,  the US-DOE
Nuclear Science Grant No. DE-SC0020633, the National Science Foundation Graduate Research Fellowship Program under Grant No. DGE-1746047, and by the U.S. Department
of Energy, Office of Science, Office of Nuclear Physics,
within the framework of the Beam Energy Scan Topical (BEST) Collaboration. 
We also acknowledge the
support from the Center of Advanced Computing and
Data Systems at the University of Houston. P.P. acknowledges support by the DFG grant SFB/TR55.

\appendix

\section{Smooth merging of correlation length}\label{sec:smoothmerg}
In order to obtain a result for the correlation length that is consistent with the scaling behavior of the 3D Ising model, we perform a smooth merging of the $\epsilon$-expansion and asymptotic forms of the correlation length, as shown in Eqs. (\ref{eq:g_epsilonexp}),~(\ref{eq:g_asymptotic}). We employ a hyperbolic tangent in order to achieve a well-behaved result, useful for hydrodynamic simulations:
\begin{equation} \label{eq:ximerging}
    \begin{split}
        \xi_{\text{final}}(T,\mu_B) = \xi_{\text{asymp}}(T,\mu_B) \frac{1}{2}
        \Big[1 \pm \tanh{\Big(\frac{T-T_{\text{low/high}}'(\mu_B)}{\Delta T(\mu_B)}}\Big)\Big] \\
        + \xi_{\epsilon}(T,\mu_B) \frac{1}{2}
        \Big[1 \mp \tanh{\Big(\frac{T-T_{\text{low/high}}'(\mu_B)}{\Delta T(\mu_B)}}\Big)\Big],
    \end{split}
\end{equation}
\begin{figure*}
    \centering
    \includegraphics[width=0.5\textwidth]{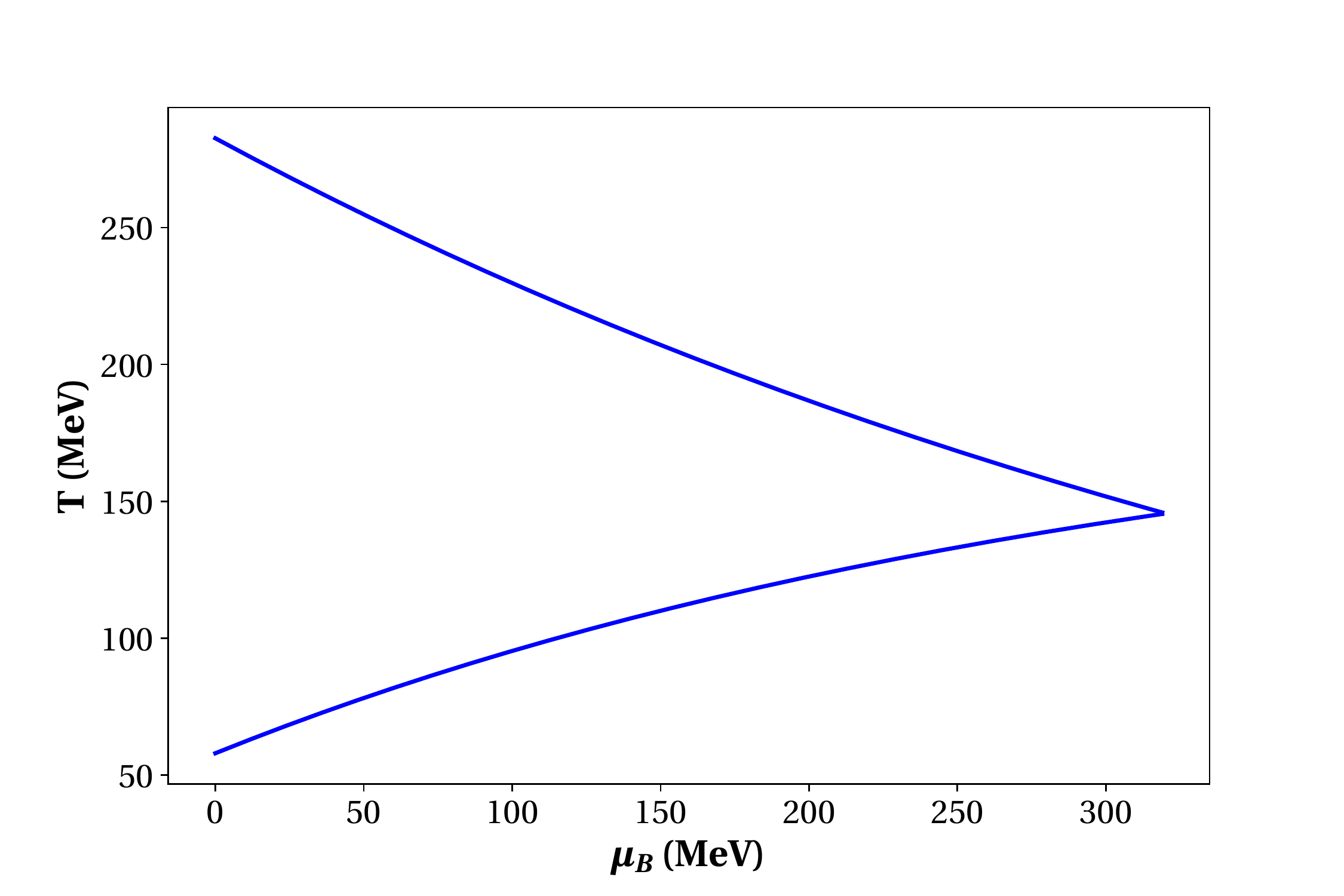}
    \caption{The curve $x=5$ in the QCD phase diagram over which the merging of the correlation length formulations is performed.}
    \label{fig:switch_for_merging}
\end{figure*}
where $T'(\mu_B)$ acts as the switching temperature and $\Delta T(\mu_B)$ is the overlap region between the two representations. The merging is performed along a curve in the $(T,\mu_B)$ plane that corresponds to a value of the scaling parameter $x=5$, as stated in Sec. \ref{sec:corr_leng}. The shape which this curve follows in the QCD phase diagram is shown in Fig. \ref{fig:switch_for_merging}. Since this is a multi-valued function of $\mu_B$, in order to fit this curve for the switching temperature, we separate it into two functions $T_{\text{low}}'(\mu_B)$ and $T_{\text{high}}'(\mu_B)$ for different temperature regimes.
These curves were best fit with exponentials, the functional forms for which are
\begin{equation} \label{eq:fitT_low}
    \begin{split}
        T_{\text{low}}’(\mu_B) = T_{\text{M,low}} - (T_{\text{M,low}} - T_{0,\text{low}}) e^{-k_{\text{low}} \mu_B}
    \end{split}
\end{equation}
\begin{equation} \label{eq:fitT_high}
    \begin{split}
        T_{\text{high}}'(\mu_B) = T_{0,\text{high}} e^{-k_{\text{high}} \mu_B},
    \end{split}
\end{equation} 
where $T_{\text{M,low}}=195$ MeV, $T_{0,\text{low}}=58$ MeV, $k_{\text{low}}=0.00318462$, $T_{0,\text{high}}=282.594$ MeV, $k_{\text{high}}=0.00207197$. We additionally write the overlap region, $\Delta T$, as a function of $\mu_B$ in order to achieve the smoothest possible result for the correlation length. For this function we also employ an exponential:
\begin{equation} \label{eq:fitT_delta}
    \begin{split}
        \Delta T(\mu_B) = T_{\text{0,merg}} + (T_{\text{M,merg}} - T_{0,\text{merg}}) e^{-k_{\text{merg}} \mu_B}
    \end{split}
\end{equation}
where $T_{\text{M,merg}}=17$ MeV, $T_{0,\text{merg}}=0.1$ MeV, $k_{\text{merg}}=0.0075$. This procedure for the smooth merging of the two representations of the correlation length through a hyperbolic tangent yields the critical contribution to the correlation length shown in Fig. \ref{fig:corrleng}. 

\bibliography{all}

\end{document}